\documentclass[twocolumn, superscriptaddress, prb]{revtex4-1}

\usepackage{hyperref}
\usepackage{amsmath,amssymb}
\usepackage{graphicx}
\usepackage{dcolumn}
\usepackage{bm}
\usepackage{multirow}
\usepackage{blindtext}
\usepackage{mathrsfs}
\usepackage{natbib}
\usepackage{braket}

\def \bsym {\boldsymbol}
\def \n {\nonumber\\}

\begin{document}
\title{Continuum model study of optical absorption by hybridized moir\'e excitons in
transition metal dichalcogenide heterobilayers}
\author{Yao-Wen Chang}
%
\affiliation{Physics Division, National Center for Theoretical Sciences, National Taiwan
University, Taipei 10617, Taiwan}

\begin{abstract}

We propose a continuum model for the theoretical study of hybridized moir\'e excitons in
transition metal dichalcogenides heterobilayers, and we use a variational method to solve
the exciton wavefunction and calculate the optical absorption spectrum. The exciton
continuum model is built by the charge continuum model for electrons and holes in moir\'e
superlattices, thereby preserving the moir\'e periodicity and lattice symmetry from the
charge continuum model. The momentum-space shift of interlayer electron-hole distribution
is included, and thus the indirect nature of interlayer excitons is described. The spin
and valley degrees of freedom and related interactions are omitted, except for the
spin-orbit energy splitting of A and B excitons. This continuum model is applied to the
simulation of optical absorption by hybridized moir\'e excitons in
$\text{WSe}_2$/$\text{WS}_2$ and $\text{MoSe}_2$/$\text{WS}_2$ heterobilayers. Twist-angle
and electric-field dependences of absorption spectra are studied. Calculated spectra are
compared with experimental observations in the literature, and correspondences of
signatures are found. The deficiency and the potential of the present model are discussed.

\end{abstract}

\maketitle

\section{Introduction}

Two-dimensional (2D) semiconducting materials, such as transition metal dichalcogenides
(TMDCs), are promising to be applied in future photonic and optoelectronic
devices\cite{wang2012electronics, xia2014two, mak2016photonics}. Excitonic effects are
critical for deciding the optical properties of 2D semiconductors due to the reduction of
dielectric screening and the enhancement of quantum confinement\cite{xiao2017excitons,
berkelbach2017optical}. Recently, moir\'e heterostructures stacked by 2D semiconducting
materials with twisted angles or lattice mismatches have been predicted theoretically and
synthesized experimentally. Emergent physical phenomena related to the additional moir\'e
degree of freedom have been observed and discussed\cite{tran2020moire, kennes2021moire,
mak2022semiconductor, zhang2023controlling, du2023moire}. Exciton physics in this system
is also studied. Excitons in moir\'e heterostructures are modulated by interlayer
charge-transfer couplings and moir\'e periodic potentials, which are formed by local
atomic registry changes\cite{kunstmann2018momentum, rivera2018interlayer,
alexeev2019resonantly, jin2019observation, jin2019identification, tran2019evidence,
deilmann2020light, shimazaki2020strongly, zhang2020twist, tang2021tuning,
jiang2021interlayer, liu2021moire, wilson2021excitons, huang2022excitons,
lin2023remarkably}. Optical properties of excitons in 2D heterostructures can be tuned by
this additional moir\'e degree of freedom. The applications of 2D layered materials are
widened by this new tunability.

An exciton is an electron-hole pair bound by Coulomb attraction\cite{xiao2017excitons,
berkelbach2017optical}. Excitonic signatures can be found in direct-gap semiconductors and
detected by optical spectroscopy. For multilayer materials, excitons can be sorted as
intralayer excitons, where the electron and the hole are at the same layer, and interlayer
excitons, where the electron and the hole are at different layers. In TMDC heterobilayers
with type-II band alignment, the lowest-energy excitation is contributed from the
interlayer exciton, which has a small electron-hole wavefunction
overlap\cite{rivera2018interlayer, deilmann2020light}. Therefore, interlayer
valence-to-conduction band transition has a low oscillator strength, and thus
interlayer-exciton signatures are difficult to be observed in absorption spectra. While
there is no direct interlayer exciton transition in TMDC heterobilayers, electron transfer
between conduction bands on each layer or hole transfer between valence bands on each
layer could occur. The charge transfer contributes to the formation of interlayer
excitons. In some situations, such as by applying a perpendicularly electric field to tune
the band gaps, the difference between the optical transition energy of interlayer excitons
and the transition energy of intralayer excitons can be at the same scale as the
interlayer charge-transfer coupling. Hybridization between interlayer excitons and
intralayer excitons forms hybrid excitons. These hybrid excitons borrow the oscillator
strength from the intralayer exciton, and thus they can be observed in absorption
spectra\cite{chaves2018electrical, hsu2019tailoring, hagel2021exciton}.

With the additional moir\'e degree of freedom in TMDC heterobilayers, excitons become
trapped by the moir\'e potential, and the translational symmetry of the center-of-mass
(COM) motion of excitons becomes broken. The interlayer charge-transfer coupling should
also follow the same moir\'e periodicity with the potential. If the physical properties of
an exciton are decided or strongly affected by this moir\'e degree of freedom, the exciton
can be called a moir\'e exciton\cite{liu2021moire, wilson2021excitons, huang2022excitons}.
Various theoretical methods have been proposed and applied to the research of moir\'e
excitons\cite{yu2015anomalous, yu2017moire, wu2017topological, wu2018theory,
danovich2018localized, ruiz2019interlayer, ruiz2020theory, brem2020hybridized,
brem2020tunable, guo2020shedding, choi2021twist, hichri2021resonance, yu2021luminescence,
julku2022nonlocal, naik2022intralayer, hagel2022electrical}. An effective continuum model
which describes the COM motion of an exciton being modulated by a periodic potential as
the moir\'e potential has been widely used to find exciton band structures and exciton
wavefunctions\cite{yu2017moire, wu2017topological, wu2018theory}. For more generalized
continuum models\cite{ruiz2019interlayer, choi2021twist, yu2021luminescence,
julku2022nonlocal}, the interlayer charge-transfer coupling is included. However, these
models only describe the COM motion of moir\'e excitons, such that the effect of moir\'e
periodicity on the internal motion of excitons has not been discussed. A more
sophisticated method is to apply atomistic calculation with density functional theory
(DFT) to simulate the hybridized moir\'e excitons\cite{brem2020hybridized,
brem2020tunable, guo2020shedding, naik2022intralayer}. However, since the unit cell for
moir\'e superlattices is usually quite large, the atomistic approach with DFT is expensive
computationally. A method to model impartially and simulate efficiently hybridized moir\'e
excitons is still lacking. It is the gap we intend to fill in this work.

In this article, we provide an improved version of the exciton continuum model to study
the hybridized moir\'e excitons in TMDC heterobilayers with lattice mismatches and small
twist angles. The exciton continuum model is written down based on four conditions. First,
the exciton continuum model is built by the charge continuum model for electrons and holes
in moir\'e superlattices, thereby preserving the moir\'e periodicity and lattice symmetry
from the charge continuum model. Second, the momentum-space shift of interlayer
electron-hole distribution is included, and thus the indirect nature of interlayer
excitons is described. Third, the spin and valley degrees of freedom and related
interactions are omitted, except for the spin-orbit energy splitting of A and B excitons.
Forth, in long moir\'e-wavelength and zero charge-transfer-coupling limits, the exciton
model and the optical absorption formula can be reduced to the counterparts of an isolated
exciton. A variational method using Slater-type orbitals (STOs) as the basis function is
applied to the numerical solution of the exciton continuum model. Optical absorption by
hybridized moir\'e excitons is studied with varying twist angles in small degrees and
varying out-of-plane electric fields to tune the band-edge energy. Twist-angle-dependent
and electric-field-dependent absorption spectra of $\text{WSe}_2$/$\text{WS}_2$ and
$\text{MoSe}_2$/$\text{WS}_2$ heterobilayers are simulated. A good correspondence between
the calculated spectra from this work and the experimentally observed spectra from the
literature is achieved. In Sec.~\ref{sec:theory}, the exciton continuum model is derived
based on the moir\'e periodicity and the charge continuum model. The methods to solve the
exciton wavefunction and the optical transition amplitude are also introduced. In
Sec.~\ref{sec:application}, the model is applied to the simulation of
twist-angle-dependent and electric-field-dependent absorption spectra of
$\text{WSe}_2$/$\text{WS}_2$ and $\text{MoSe}_2$/$\text{WS}_2$ heterobilayers. In
Sec.~\ref{sec:discussion}, the applications and improvements of this model in the future
study are discussed, and the conclusion is given. In Appendix~\ref{sec:matrix_element},
matrix elements of the exciton continuum model spanned by STO and plane-wave basis
functions are given.

\section{Theory\label{sec:theory}}

In this section, the continuum model for hybridized moir\'e excitons in TMDC
heterobilayers is derived, and the method to calculate the exciton wavefunction and the
optical spectrum is given. In Sec.~\ref{ssec:superlattice}, the geometry of moir\'e
superlattices and reciprocal lattices is illustrated, and the concept of moir\'e Brillouin
zone (MBZ) is introduced. In Sec.~\ref{ssec:charge_model}, the charge continuum model and
the many-body formulation are introduced. The expression of screened Coulomb interaction
in the many-body Hamiltonian is also discussed. In Sec.~\ref{ssec:exciton_model}, the
exciton continuum model is derived through the many-body formulation. A transformation
from electron and hole position coordinates to exciton COM and internal coordinates is
applied to the continuum model. In Sec.~\ref{ssec:exciton_wavefunction}, the variational
method is introduced to solve the exciton wavefunction. In
Sec.~\ref{ssec:optical_absorption}, the method to calculate the optical absorption
spectrum is discussed.

\subsection{Moir\'e superlattice and MBZ\label{ssec:superlattice}}

\begin{figure}
\includegraphics[width=0.95\linewidth]{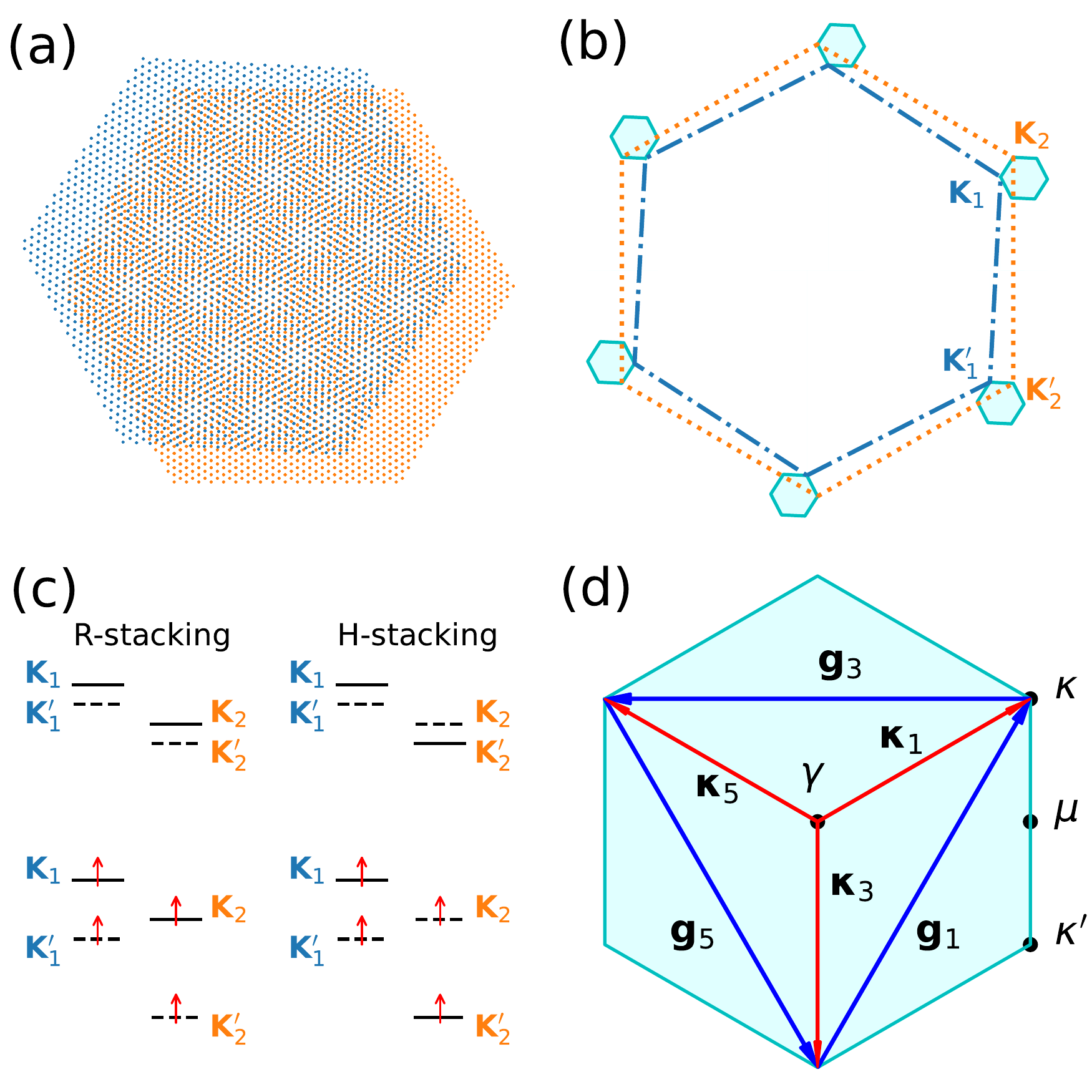}
\caption{(a) Schematic plot of the R-stacked heterobilayer with lattice mismatch
$\delta=0.1$ and a twisted angle $\theta=3^{\circ}$. (b) Schematic plot of the Brillouin
zone of the top monolayer (inside blue dash-dot hexagonal) and the Brillouin zone of the
bottom monolayer (inside orange dot hexagonal) of the R-stacked heterobilayer. The filled
zone is the MBZ of heterobilayers. (c) Energy-level diagram of band edges at the top
monolayer $\mathbf{K}_1$, $\mathbf{K}'_1$ and band edges at the bottom monolayer
$\mathbf{K}_2$, $\mathbf{K}'_2$ in R-stacked and H-stacked heterobilayers. (d) A closer
look at the MBZ of the heterobilayer and high-symmetry points.} 
\label{fig:moire_BZ} 
\end{figure}

There are two types of stacking for TMDC heterobilayers: R-stacking and H-stacking.
R-stacked TMDC heterobilayers are stacking of two triangular lattices with near
$0^{\circ}$ rotation and H-stacked TMDC heterobilayers are stacking with near $60^{\circ}$
rotation\cite{jiang2021interlayer}. Small twist-angle variants straying from $0^{\circ}$
and $60^{\circ}$ rotations can still be considered as R-stacked and H-stacked
heterobilayers. Schematic plots of the moir\'e superlattice and MBZ of the R-stacked
heterobilayer are illustrated in Fig.~\ref{fig:moire_BZ} (a), (b). The lattice constant of
the moir\'e superlattice $a_{\text{M}}$ as a function of twist angle $\theta$ and the
lattice mismatch $\delta=|a'_0-a_0|/a_0$, with $a_0$, $a'_0$ the lattice constants for the
atomistic lattices, is given by\cite{tran2020moire, zhang2023controlling}
\begin{eqnarray}
  a_{\text{M}}=\frac{(1+\delta)a_0}{\sqrt{2(1+\delta)(1-\cos\theta)+\delta^2}}.
\end{eqnarray}
The MBZ is built by the difference between reciprocal vectors in the Brillouin zone of the
top monolayer (inside blue dash-dot hexagonal) and the Brillouin zone of the bottom
monolayer (inside orange dot hexagonal) as illustrated in Fig.~\ref{fig:moire_BZ} (b).
The moir\'e superlattice can be considered a triangular lattice with the unit cell
containing local geometry. The primitive vectors of the moir\'e superlattice can be
defined as $\bsym{a}_1 =a_{\text{M}}[(\sqrt{3}/2)\mathbf{e}_{x}+(1/2)\mathbf{e}_{y}]$,
$\bsym{a}_2=a_{\text{M}}[(-\sqrt{3}/2)\mathbf{e}_{x}+(1/2)\mathbf{e}_{y}]$. For the
moir\'e reciprocal lattice, the reciprocal primitive vectors can be defined by
$\bsym{g}_{i}\cdot\bsym{a}_{j}=2\pi\delta_{ij}$ for $i,j=1,2$ and found to be
$\bsym{g}_{1}=\sqrt{3}k_{\text{M}}[(1/2)\mathbf{e}_{x`} +(\sqrt{3}/2)\mathbf{e}_{y}]$,
$\bsym{g}_{2} =\sqrt{3}k_{\text{M}}[(-1/2)\mathbf{e}_{x} +(\sqrt{3}/2)\mathbf{e}_{y}]$,
with $k_{\text{M}} =4\pi/(3a_{\text{M}})$. A set of reciprocal primitive vectors can be
defined as $\bsym{g}_{j} = \sqrt{3}k_{\text{M}}\left[\mathbf{e}_{x}\cos({j\pi}/{3})
+\mathbf{e}_{y}\sin({j\pi}/{3})\right]$, with $j=1,2,\cdots,6$. The vector connects
between $\kappa$ and $\gamma$ is given by $\bsym{\kappa}_1=\left(2\bsym{g}_{1}
-\bsym{g}_{2}\right)/3$ and the vector connects between $\kappa'$ and $\gamma$ is given by
$\bsym{\kappa}_2=\left(\bsym{g}_{1}-2\bsym{g}_{2}\right)/3$. These two vectors
$\bsym{\kappa}_1$ and $\bsym{\kappa}_2$ can be assigned as reciprocal basis vectors. A set
of reciprocal basis vectors can be defined as $\bsym{\kappa}_{j} =
k_{\text{M}}\left[\mathbf{e}_{x}\cos(\pi/6-{j\pi}/{3})
+\mathbf{e}_{y}\sin(\pi/6-{j\pi}/{3})\right]$, with $j=1,2,\cdots,6$. Part of the
reciprocal primitive and basis vectors are shown in Fig.~\ref{fig:moire_BZ} (d).

Energy-level diagrams of band edges on different layers with the same spin index on
R-stacked and H-stacked heterobilayers are illustrated in Fig.~\ref{fig:moire_BZ}
(c)\cite{rivera2018interlayer}. Spin-orbit coupling (SOC) induced energy level splittings
occur in both conduction bands and valence bands of both layers. Due to the intrinsic
spin-valley locking of TMDCs, the $\mathbf{K}$ and $\mathbf{K}'$ valleys with the same
spin index have different band-edge energies. As illustrated in Fig.~\ref{fig:moire_BZ}
(b), interlayer tunneling only occurs among energy levels with the same types of lines
(solid line or dash line). Optical transition is only allowed between valence and
conduction bands at the same valley ($\mathbf{K}_1$, $\mathbf{K}'_1$, $\mathbf{K}_2$,
$\mathbf{K}'_2$). Via this diagram and effective-mass-model parameters for TMDC
monolayers, band-edge energies for different intralayer or interlayer excitons in
R-stacked or H-stacked heterobilayers can be evaluated.

\subsection{Charge continuum model\label{ssec:charge_model}}

The many-body Hamiltonian for electrons and holes in moir\'e superlattices including the
electron-hole attraction is given by
\begin{eqnarray}
  \hat{\mathcal{H}}
  &=&
  \sum_{\mathbf{k}\mathbf{k}',ll'}\Big[\hat{c}^{\dagger}_{l,\mathbf{k}}
  \tilde{H}_{\text{e},ll'}(\mathbf{k},\mathbf{k}')\hat{c}_{l',\mathbf{k}'}
  +\hat{d}^{\dagger}_{l,\mathbf{k}}
  \tilde{H}_{\text{h},ll'}(\mathbf{k},\mathbf{k}')\hat{d}_{l',\mathbf{k}'}\Big]\n
  &&+\frac{1}{2S}\sum_{ll',\mathbf{q}}\tilde{W}_{ll'}(\mathbf{q})
  \Big(\hat{\varrho}_{\text{e},l,\mathbf{q}}\hat{\varrho}_{\text{e},l',-\mathbf{q}}
  +\hat{\varrho}_{\text{h},l,\mathbf{q}}\hat{\varrho}_{\text{h},l',-\mathbf{q}}\Big)\n
  &&-\frac{1}{S}\sum_{ll',\mathbf{q}}\tilde{W}_{ll'}(\mathbf{q})
  \hat{\varrho}_{\text{e},l,\mathbf{q}}\hat{\varrho}_{\text{h},l',-\mathbf{q}},
\end{eqnarray}
where $\hat{c}^{\dagger}_{l,\mathbf{k}}$/$\hat{c}_{l,\mathbf{k}}$ is the electron
creation/annihilation operators at layer $l$ with wavevector $\mathbf{k}$,
$\hat{d}^{\dagger}_{l,\mathbf{k}}$/$\hat{d}_{l,\mathbf{k}}$ is the hole
creation/annihilation operators, $\tilde{H}_{\text{c},ll'}(\mathbf{k},\mathbf{k}')$ is the
charge Hamiltonian with $\text{c}=\{\text{e},\text{h}\}$ the charge index indicating
electrons and holes, $\hat{\varrho}_{\text{e},l,\mathbf{q}} = \sum_{\mathbf{k}}
\hat{c}^{\dagger}_{l,\mathbf{k}+\mathbf{q}}\hat{c}_{l,\mathbf{k}}$ is the electron density
operator, $\hat{\varrho}_{\text{h},l,\mathbf{q}} = \sum_{\mathbf{k}}
\hat{d}^{\dagger}_{l,\mathbf{k}+\mathbf{q}}\hat{d}_{l,\mathbf{k}}$ is the hole density
operator, $S$ is the area of the lattice, and $\tilde{W}_{ll'}(\mathbf{q}) =\int
e^{-\mathtt{i}\mathbf{q}\cdot\mathbf{r}} {W}_{ll'}(\mathbf{r})\text{d}^2r$ is the screened
Coulomb interaction. Note that we have ignored the spin and valley degrees of freedom in
this many-body Hamiltonian.

The charge Hamiltonian in real space is related to the charge Hamiltonian in $k$-space
through the Fourier transform $\tilde{H}_{\text{c},ll'}(\mathbf{k},\mathbf{k}') = \int
e^{-\mathtt{i}(\mathbf{k}-\mathbf{k}')\cdot\mathbf{r}}
{H}_{\text{c},ll'}(\mathbf{r})\text{d}^2r$ and is given by\cite{wu2018hubbard}
\begin{eqnarray}
  H_{\text{c}}(\mathbf{r})
  =
  \begin{pmatrix}
    h_{\text{c},1}(\mathbf{r}) & t_{\text{c}}(\mathbf{r})\\
    t^*_{\text{c}}(\mathbf{r}) & h_{\text{c},2}(\mathbf{r})
  \end{pmatrix},
\end{eqnarray}
where $h_{\text{c},l}(\mathbf{r})$ is the layer-charge Hamiltonian of the $l$-th layer and
$t_{\text{e}}(\mathbf{r})$, $t_{\text{h}}(\mathbf{r})$ is the interlayer tunneling
coupling. The layer-charge Hamiltonian is given by
\begin{eqnarray}
  h_{\text{c},l}(\mathbf{r})
  &=&
  \epsilon_{\text{c},l}-z_{\text{c}}U_{l}(\mathbf{r})
  +\frac{|\mathbf{p}+z_{\text{c}}\bsym{\kappa}_{l}|^2}{2m_{\text{c},l}},
\end{eqnarray}
where $\mathbf{p}=-\mathtt{i}\bsym{\nabla}$ is the momentum operator of the charge
particle, $z_{\text{e}}=-$ and $z_{\text{h}}=+$ are the charge parity,
$\epsilon_{\text{c},l}$ is the band-edge energy, $m_{\text{c},l}$ is the charge mass at
layer $l$, $U_{l}(\mathbf{r})$ is the moir\'e potential. The moir\'e potential is given by
the harmonic-function form
\begin{eqnarray}
  U_{l}(\mathbf{r})
  =
  2V_{l}\sum_{j=1,3,5}\cos\left(\bsym{g}_{j}\cdot\mathbf{r}+\psi_{l}\right),
\end{eqnarray}
and the interlayer tunneling couplings are given by
\begin{eqnarray}
  t_{\text{e}}(\mathbf{r})
  =
  w_{\text{e}}\left(1+e^{\mathtt{i}\bsym{g}_{1}\cdot\mathbf{r}}
  +e^{\mathtt{i}\bsym{g}_{2}\cdot\mathbf{r}}\right),
\end{eqnarray}
\begin{eqnarray}
  t_{\text{h}}(\mathbf{r})
  =
  w_{\text{h}}\left(1+e^{-\mathtt{i}\bsym{g}_{1}\cdot\mathbf{r}}
  +e^{-\mathtt{i}\bsym{g}_{2}\cdot\mathbf{r}}\right),
\end{eqnarray}
where $V_{l}$ is the moir\'{e} potential depth at layer $l$, $\psi_{l}$ is the phase
angle, and $w_{\text{e}}$/$w_{\text{h}}$ are the electron/hole interlayer-tunneling
coupling strength. In this work, the phase angle is always chosen to be
$\psi_{l}=-(-1)^{l}\pi/2$, such that the moir\'e potential can be rewritten as
$U_{l}(\mathbf{r}) = 2V_{l}(-1)^{l}\sum_{j=1,3,5}
\sin\left(\bsym{g}_{j}\cdot\mathbf{r}\right)$.

The screened Coulomb potential is described by a modified Rytova-Keldysh
potential\cite{rytova, keldysh} which is written as
\begin{eqnarray}
  \tilde{W}_{ll'}(\mathbf{q})= \frac{2\pi}{\epsilon_{ll'}({q})q},
\end{eqnarray}
where $\epsilon_{ll'}({q})$ is the bilayer dielectric function determined by intrinsic
properties of heterobilayers. While there are many formulations for the bilayer dielectric
function\cite{danovich2018localized, ruiz2020theory, thygesen2017calculating,
cavalcante2018electrostatics, van2018interlayer, florez2020effects}, here we use a simple
form derived in Appendix~\ref{sec:RK_potential}. The intralayer and interlayer dielectric
functions are given by
\begin{eqnarray}
  \epsilon_{11}(q)
  &=&
  \frac{\kappa_{0}\epsilon_{12}(q)}
  {\left(\frac{\kappa_{1}+\kappa_{0}}{2}+r_{2}q\right)e^{q{d}}
  -\left(\frac{\kappa_{1}-\kappa_{0}}{2}+r_{2}q\right)e^{-q{d}}},
  \label{intra_1}
\end{eqnarray}
\begin{eqnarray}
  \epsilon_{22}(q)
  &=&
  \frac{\kappa_{0}\epsilon_{12}(q)}
  {\left(\frac{\kappa_{1}+\kappa_{0}}{2}+r_{1}q\right)e^{q{d}}
  -\left(\frac{\kappa_{1}-\kappa_{0}}{2}+r_{1}q\right)e^{-q{d}}},\hskip1ex
  \label{intra_2}
\end{eqnarray}
\begin{eqnarray}
  \epsilon_{12}(q)
  &=&
  \left(\frac{\kappa_{1}+\kappa_{0}}{2}+r_{1}q\right)
  \left(\frac{\kappa_{1}+\kappa_{0}}{2}+r_{2}q\right)\frac{e^{q{d}}}{\kappa_{0}}\n
  &&-
  \left(\frac{\kappa_{1}-\kappa_{0}}{2}+r_{1}q\right)
  \left(\frac{\kappa_{1}-\kappa_{0}}{2}+r_{2}q\right)\frac{e^{-q{d}}}{\kappa_{0}},\n
  \label{inter}
\end{eqnarray}
and $\epsilon_{21}(q)=\epsilon_{12}(q)$, where $\kappa_{1}$ is the dielectric constant
outside the bilayer, $\kappa_{0}$ is the dielectric constant inside the bilayer, $r_{l}$
is the screening length on the $l$-th layer, and $d$ is the interlayer distance.

\subsection{Exciton continuum model\label{ssec:exciton_model}}

An exciton state is assumed to be written as
$\ket{\text{X}_{I}}=\text{X}^{\dagger}_{I}\ket{0}$, where
\begin{eqnarray}
  \text{X}^{\dagger}_{I}
  &=&
  \sum_{l_{\text{e}},l_{\text{h}}}\sum_{\mathbf{k}_{\text{e}},\mathbf{k}_{\text{h}}}
  \tilde{\Psi}_{l_{\text{e}}l_{\text{h}},I}(\mathbf{k}_{\text{e}},\mathbf{k}_{\text{h}})
  \hat{c}^{\dagger}_{l_{\text{e}},\mathbf{k}_{\text{e}}}
  \hat{d}^{\dagger}_{l_{\text{h}},\mathbf{k}_{\text{h}}}
\end{eqnarray}
is the exciton creation operator, $\ket{0}$ is the ground-state ket, and
$\tilde{\Psi}_{I}(\mathbf{k}_{\text{e}},\mathbf{k}_{\text{h}})$ is the exciton
wavefunction. By variation method, $\delta[\braket{\text{X}_{I}|\hat{\mathcal{H}}
|\text{X}_{I}} -\lambda(\braket{\text{X}_{I}|\text{X}_{I}}-1)] = 0$, an eigenvalue
equation can be derived as
\begin{eqnarray}
  \sum_{\mathbf{k}'_{\text{e}},\mathbf{k}'_{\text{h}}}
  \tilde{\mathcal{H}}_{\text{X}}(\mathbf{k}_{\text{e}},\mathbf{k}_{\text{h}};
  \mathbf{k}'_{\text{e}},\mathbf{k}'_{\text{h}})
  \tilde{\Psi}_{I}(\mathbf{k}'_{\text{e}},\mathbf{k}'_{\text{h}})
  =
  \varepsilon_{\text{X},I}\tilde{\Psi}_{I}(\mathbf{k}_{\text{e}},\mathbf{k}_{\text{h}}),\n
\end{eqnarray}
where $\tilde{\mathcal{H}}_{\text{X}}(\mathbf{k}_{\text{e}},\mathbf{k}_{\text{h}};
\mathbf{k}'_{\text{e}},\mathbf{k}'_{\text{h}}) = \bra{0}
\hat{d}_{l_{\text{h}},\mathbf{k}_{\text{h}}} \hat{c}_{l_{\text{e}},\mathbf{k}_{\text{e}}}
\hat{\mathcal{H}} \hat{c}^{\dagger}_{l'_{\text{e}},\mathbf{k}'_{\text{e}}}
\hat{d}^{\dagger}_{l'_{\text{h}},\mathbf{k}'_{\text{h}}} \ket{0}$ is the momentum-space
exciton Hamiltonian, and $\varepsilon_{\text{X},I}$ is the exciton eigenenergy. The
continuum model to describe excitons in 2D moir\'{e} superlattices can be derived from the
Fourier transform of the momentum-space exciton Hamiltonian
\begin{eqnarray}
  \mathcal{H}_{\text{X}}(\mathbf{r}_{\text{e}},\mathbf{r}_{\text{h}})
  &=&
  \frac{1}{N}\sum_{\Delta\mathbf{k}_{\text{e}},\Delta\mathbf{k}_{\text{h}}}
  \exp\left({\mathtt{i}\Delta\mathbf{k}_{\text{e}}\cdot\mathbf{r}_{\text{e}}}\right)
  \exp\left({\mathtt{i}\Delta\mathbf{k}_{\text{h}}\cdot\mathbf{r}_{\text{h}}}\right)\n
  &&\times\tilde{\mathcal{H}}_{\text{X}}
  (\mathbf{k}_{\text{e}}+\Delta\mathbf{k}_{\text{e}},
  \mathbf{k}_{\text{h}}+\Delta\mathbf{k}_{\text{h}};
  \mathbf{k}_{\text{e}},\mathbf{k}_{\text{h}}),
\end{eqnarray}
and it is written as
\begin{eqnarray}
  &&\mathcal{H}_{\text{X}}(\mathbf{r}_{\text{e}},\mathbf{r}_{\text{h}})=\n
  &&
  \begin{pmatrix}
    \mathcal{H}_{11}(\mathbf{r}_{\text{e}},\mathbf{r}_{\text{h}}) & 0
    & \mathcal{T}_{\text{h}}(\mathbf{r}_{\text{e}},\mathbf{r}_{\text{h}})
    & \mathcal{T}_{\text{e}}(\mathbf{r}_{\text{e}},\mathbf{r}_{\text{h}})\\
    0 & \mathcal{H}_{22}(\mathbf{r}_{\text{e}},\mathbf{r}_{\text{h}})
    & \mathcal{T}^*_{\text{e}}(\mathbf{r}_{\text{e}},\mathbf{r}_{\text{h}})
    & \mathcal{T}^*_{\text{h}}(\mathbf{r}_{\text{e}},\mathbf{r}_{\text{h}})\\
    \mathcal{T}^*_{\text{h}}(\mathbf{r}_{\text{e}},\mathbf{r}_{\text{h}})
    & \mathcal{T}_{\text{e}}(\mathbf{r}_{\text{e}},\mathbf{r}_{\text{h}})
    & \mathcal{H}_{12}(\mathbf{r}_{\text{e}},\mathbf{r}_{\text{h}}) & 0\\
    \mathcal{T}^*_{\text{e}}(\mathbf{r}_{\text{e}},\mathbf{r}_{\text{h}})
    & \mathcal{T}_{\text{h}}(\mathbf{r}_{\text{e}},\mathbf{r}_{\text{h}}) &
    0 & \mathcal{H}_{21}(\mathbf{r}_{\text{e}},\mathbf{r}_{\text{h}})
  \end{pmatrix},\n
\end{eqnarray}
where $\mathcal{T}_{\text{e}}(\mathbf{r}_{\text{e}},\mathbf{r}_{\text{h}})
=t_{\text{e}}(\mathbf{r}_{\text{e}})$,
$\mathcal{T}_{\text{h}}(\mathbf{r}_{\text{e}},\mathbf{r}_{\text{h}})
=t_{\text{h}}(\mathbf{r}_{\text{e}})$ are interlayer charge-transfer couplings,
\begin{eqnarray}
  \mathcal{H}_{l_{\text{e}}l_{\text{h}}}(\mathbf{r}_{\text{e}},\mathbf{r}_{\text{h}})
  &=&
  \Delta_{l_{\text{e}}l_{\text{h}}}
  +\frac{|\mathbf{p}_{\text{e}}-\bsym{\kappa}_{l_\text{e}}|^2}{2m_{\text{e},l_\text{e}}}
  +\frac{|\mathbf{p}_{\text{h}}+\bsym{\kappa}_{l_\text{h}}|^2}{2m_{\text{h},l_\text{h}}}\n
  &&+U_{l_\text{e}}(\mathbf{r}_{\text{e}})-U_{l_\text{h}}(\mathbf{r}_{\text{h}})
  -W_{l_{\text{e}}l_{\text{h}}}(\mathbf{r}_{\text{eh}}),
\end{eqnarray}
is the layer-exciton Hamiltonian, $\Delta_{l_{\text{e}}l_{\text{h}}}
=\epsilon_{\text{e},l_{\text{e}}} +\epsilon_{\text{h},l_{\text{h}}}$ is the band-gap
energy, and $W_{l_{\text{e}}l_{\text{h}}}(\mathbf{r})=\int
e^{\mathtt{i}\mathbf{k}\cdot\mathbf{r}}\tilde{W}(\mathbf{k})\text{d}^2k/(2\pi)^2$ the
screened Coulomb potential. The exciton continuum model can be reformulated by the
coordinate transformation from electron position and hole position coordinates to exciton
COM and internal coordinates, $\mathbf{R} = \gamma_{\text{e},l_\text{e}}
\mathbf{r}_{\text{e}} + \gamma_{\text{h},l_\text{h}} \mathbf{r}_{\text{h}}$, $\mathbf{r} =
\mathbf{r}_{\text{e}} -\mathbf{r}_{\text{h}}$, with $\gamma_{\text{e},l_\text{e}} =
m_{\text{e},l_\text{e}} /(m_{\text{e},l_\text{e}} +m_{\text{h},l_\text{h}})$,
$\gamma_{\text{h},l_\text{h}} = m_{\text{h},l_\text{h}} /(m_{\text{e},l_\text{e}} +
m_{\text{h},l_\text{h}})$, which leads to the following transformation for momentums
$\mathbf{p}_{\text{e}} = \gamma_{\text{e},l_\text{e}} \mathbf{P} +\mathbf{p}$,
$\mathbf{p}_{\text{h}} = \gamma_{\text{h},l_\text{h}} \mathbf{P} -\mathbf{p}$, with
$\mathbf{P}$ the COM momentum and $\mathbf{p}$ the internal momentum. The same method has
been used in Ref.~\onlinecite{hannachi2023moire}. The reciprocal basis vectors are also
applied by the coordinate transformation
\begin{eqnarray}
  \bsym{\mathcal{K}}_{l_\text{e}l_\text{h}}
  =
  \bsym{\kappa}_{l_\text{e}}-\bsym{\kappa}_{l_\text{h}},
  \hskip2ex
  \bsym{\kappa}_{l_\text{e}l_\text{h}}
  =
  \gamma_{\text{h},l_\text{h}}\bsym{\kappa}_{l_\text{e}}
  +\gamma_{\text{e},l_\text{e}}\bsym{\kappa}_{l_\text{h}},
\end{eqnarray}
with $\bsym{\mathcal{K}}_{l_\text{e}l_\text{h}}$ the COM reciprocal basis vector and
$\bsym{\kappa}_{l_\text{e}l_\text{h}}$ the internal reciprocal basis vector. The exciton
Hamiltonian can be rewritten as
\begin{eqnarray}
  &&\mathcal{H}_{\text{X}}(\mathbf{R},\mathbf{r})\n
  &&=
  \begin{pmatrix}
    \mathcal{H}_{11}(\mathbf{R},\mathbf{r}) & 0
    & \mathcal{T}_{\text{h}}(\mathbf{R},\mathbf{r})
    & \mathcal{T}_{\text{e}}(\mathbf{R},\mathbf{r})\\
    0 & \mathcal{H}_{22}(\mathbf{R},\mathbf{r})
    & \mathcal{T}^*_{\text{e}}(\mathbf{R},\mathbf{r})
    & \mathcal{T}^*_{\text{h}}(\mathbf{R},\mathbf{r})\\
    \mathcal{T}^*_{\text{h}}(\mathbf{R},\mathbf{r})
    & \mathcal{T}_{\text{e}}(\mathbf{R},\mathbf{r})
    & \mathcal{H}_{12}(\mathbf{R},\mathbf{r}) & 0\\
    \mathcal{T}^*_{\text{e}}(\mathbf{R},\mathbf{r})
    & \mathcal{T}_{\text{h}}(\mathbf{R},\mathbf{r}) & 0
    & \mathcal{H}_{21}(\mathbf{R},\mathbf{r})
  \end{pmatrix},\n
\end{eqnarray}
where
\begin{eqnarray}
  \mathcal{T}_{\text{e}}(\mathbf{R},\mathbf{r})
  &=&
  w_{\text{e}}\big[1+\exp\left({\mathtt{i}\bsym{g}_{1}\cdot
  \left(\mathbf{R}+\gamma_{\text{h},l_{\text{h}}}\mathbf{r}\right)}\right)\n
  &&+\exp\left({\mathtt{i}\bsym{g}_{2}\cdot
  \left(\mathbf{R}+\gamma_{\text{h},l_{\text{h}}}\mathbf{r}\right)}\right)\big],
\end{eqnarray}
\begin{eqnarray}
  \mathcal{T}_{\text{h}}(\mathbf{R},\mathbf{r})
  &=&
  w_{\text{h}}\big[1+\exp\left(-{\mathtt{i}\bsym{g}_{1}\cdot
  \left(\mathbf{R}-\gamma_{\text{e},l_{\text{e}}}\mathbf{r}\right)}\right)\n
  &&+\exp\left(-{\mathtt{i}\bsym{g}_{2}\cdot
  \left(\mathbf{R}-\gamma_{\text{e},l_{\text{e}}}\mathbf{r}\right)}\right)\big],
\end{eqnarray}
\begin{eqnarray}
  \mathcal{H}_{l_{\text{e}}l_{\text{h}}}(\mathbf{R},\mathbf{r})
  &=&
  \Delta_{l_{\text{e}}l_{\text{h}}}
  +\frac{|\mathbf{P}-\bsym{\mathcal{K}}_{l_\text{e}l_\text{h}}|^2}
  {2m_{\text{X},l_{\text{e}}l_{\text{h}}}}
  +\frac{|\mathbf{p}-\bsym{\kappa}_{l_\text{e}l_\text{h}}|^2}
  {2\mu_{\text{X},l_{\text{e}}l_{\text{h}}}}\n
  &&+\mathcal{V}_{l_{\text{e}}l_{\text{h}}}(\mathbf{R},\mathbf{r})
  -W_{l_{\text{e}}l_{\text{h}}}({r}),
\end{eqnarray}
with $m_{\text{X},l_{\text{e}}l_{\text{h}}}$ the exciton mass,
$\mu_{\text{X},l_{\text{e}}l_{\text{h}}} $ the reduced mass, and
$\mathcal{V}_{l_{\text{e}}l_{\text{h}}}(\mathbf{R},\mathbf{r})$ the exciton moir\'e
potential. The exciton mass and the reduced mass are given by
\begin{eqnarray} 
  m_{\text{X},l_{\text{e}}l_{\text{h}}} 
  =
  m_{\text{e},l_\text{e}} + m_{\text{h},l_\text{h}},\hskip2ex
  \mu_{\text{X},l_{\text{e}}l_{\text{h}}} 
  =
  \frac{m_{\text{e},l_\text{e}}m_{\text{h},l_\text{h}}}
  {m_{\text{X},l_{\text{e}}l_{\text{h}}}}.
\end{eqnarray}
The exciton moir\'e potential is a combination of the electron moir\'e potential and the
hole moir\'e potential. The exciton moir\'e potential is given by
\begin{eqnarray}
  \mathcal{V}_{l_{\text{e}}l_{\text{h}}}(\mathbf{R},\mathbf{r})
  &=&
  U_{l_{\text{e}}}(\mathbf{R}+\gamma_{\text{h},l_{\text{h}}}\mathbf{r})
  -U_{l_{\text{h}}}(\mathbf{R}-\gamma_{\text{e},l_{\text{e}}}\mathbf{r}).
\end{eqnarray}
Note that the exciton band minimum is centered at
$\bsym{\mathcal{K}}_{l_\text{e}l_\text{h}} = \bsym{\kappa}_{l_\text{e}}
-\bsym{\kappa}_{l_\text{h}}$. For intralayer excitons $l_\text{e}=l_\text{h}=l$ thus
$\bsym{\mathcal{K}}_{l_\text{e}l_\text{h}}=0$, the exciton band minimum locates at
$\gamma$ point in MBZ. For interlayer exciton, $l_\text{e}=1$, $l_\text{h}=2$ or
$l_\text{e}=2$, $l_\text{h}=l$ thus $\bsym{\mathcal{K}}_{l_\text{e}l_\text{h}}
=\pm(\bsym{\kappa}_{2}-\bsym{\kappa}_{1})$, the exciton band minimum locates at one of
$\kappa$, $\kappa'$ points in the MBZ. Based on Fig.~\ref{fig:moire_BZ} (b), the momentum
shift of interlayer excitons comes from the momentum difference between $\mathbf{K}_1$
($\mathbf{K}'_1$) and $\mathbf{K}_2$ ($\mathbf{K}'_2$) points on the Brillouin zones of
the top and bottom layers. Therefore, the intralayer exciton is assigned as a direct
exciton and the interlayer exciton is assigned as an indirect exciton. Optical transitions
only occur for intralayer excitons located at $\mathbf{K}=\mathbf{0}$, which is $\gamma$
point on the MBZ of excitons.

It is worth a mention that, by Taylor expansion of the internal coordinate, the exciton
moir\'e potential can be approximated as
\begin{eqnarray}
  \mathcal{V}_{l_{\text{e}}l_{\text{h}}}(\mathbf{R},\mathbf{r})
  &\simeq&
  U_{l_{\text{e}}}(\mathbf{R})
  +\gamma_{\text{h},l_{\text{h}}}\mathbf{r}\cdot\bsym{\nabla}_{\mathbf{R}}
  U_{l_{\text{e}}}(\mathbf{R})\n
  &&-U_{l_{\text{h}}}(\mathbf{R})
  +\gamma_{\text{e},l_{\text{e}}}\mathbf{r}\cdot\bsym{\nabla}_{\mathbf{R}}
  U_{l_{\text{h}}}(\mathbf{R}).
\end{eqnarray}
For intralayer excitons, $l_{\text{e}}=l_{\text{h}}=l$ and $\psi_l=-(-1)^l\pi/2$ are used,
and the exciton moir\'e potential becomes
\begin{eqnarray}
  \mathcal{V}_{ll}(\mathbf{R},\mathbf{r})
  &\simeq&
  2{V}_{l}(-1)^{l}\sum_{j=1,3,5}
  \left(\bsym{g}_{j}\cdot\mathbf{r}\right)\cos(\bsym{g}_{j}\cdot\mathbf{R}).
\end{eqnarray}
On the other hand, for interlayer excitons, $l_{\text{e}}=2$ and $l_{\text{h}}=1$, the
exciton moir\'e potential becomes
\begin{eqnarray}
  \mathcal{V}_{21}(\mathbf{R},\mathbf{r})
  &\simeq&
  2({V}_{1}+V_{2})\sum_{j=1,3,5}\sin(\bsym{g}_{j}\cdot\mathbf{R}).
\end{eqnarray}
Therefore, exciton moir\'e potentials for intralayer excitons and interlayer excitons are
different. Additionally, the exciton moir\'e potential jointly couples the COM motion and
the internal motion. The COM wavefunction and the internal wavefunction of an intralayer
moir\'e exciton are entangled for any moir\'e potential depths.

\subsection{Exciton wavefunction\label{ssec:exciton_wavefunction}}

The exciton wavefunction and exciton band structure can be solved by the eigenvalue problem
\begin{eqnarray}
  \mathcal{H}_{\text{X}}(\mathbf{R},\mathbf{r})
  \Psi_{\text{X},I\mathbf{K}}(\mathbf{R},\mathbf{r})
  =
  \varepsilon_{\text{X},I\mathbf{K}}
  \Psi_{\text{X},I\mathbf{K}}(\mathbf{R},\mathbf{r}),
\end{eqnarray}
where $\Psi_{\text{X},I\mathbf{K}}(\mathbf{R},\mathbf{r})$ is the exciton wavefunction and
$\varepsilon_{\text{X},I\mathbf{K}}$ is the exciton eigenenergy. A variational exciton
wavefunction method is used to solve the exciton Hamiltonian. To include the entanglement
between the COM motion and the internal motion, the variational exciton wavefunction
should contain a direct product of the COM-coordinate function and the internal-coordinate
function. Such variational exciton wavefunction is written as
\begin{eqnarray}
  \Psi_{\text{X},I\mathbf{K}}(\mathbf{R},\mathbf{r})
  &=&
  \frac{e^{\mathtt{i}\mathbf{K}\cdot\mathbf{R}}}{2\pi}
  \sum_{l_{\text{e}}l_{\text{h}},a,\mathbf{G}}
  C_{(l_{\text{e}}l_{\text{h}},{a},\mathbf{G}),I\mathbf{K}}\;
  \chi_{\mathbf{G}}(\mathbf{R})\n
  &&\times
  \exp\left({\mathtt{i}\bsym{\kappa}_{l_\text{e}l_\text{h}}\cdot\mathbf{r}}\right)
  \phi_{l_{\text{e}}l_{\text{h}},a}(\mathbf{r})\mathbb{E}_{l_{\text{e}}l_{\text{h}}},
  \label{exciton_wavefn}
\end{eqnarray}
where $C_{(l_{\text{e}}l_{\text{h}},{a},\mathbf{G}),I\mathbf{K}}$ is the variational
wavefunction coefficient, $\chi_{\mathbf{G}}(\mathbf{R})$ is the COM-coordinate basis
function, $\phi_{l_{\text{e}}l_{\text{h}},a}(\mathbf{r})$ is the internal-coordinate basis
function, and $\mathbb{E}_{l_{\text{e}}l_{\text{h}}}$ is the exciton Hamiltonian-matrix
basis vector with
\begin{eqnarray}
  \mathbb{E}_{11}
  =
  \begin{pmatrix}
    1\\ 0\\ 0\\ 0
  \end{pmatrix},\hskip1ex
  \mathbb{E}_{22}
  =
  \begin{pmatrix}
    0\\ 1\\ 0\\ 0
  \end{pmatrix},\hskip1ex
  \mathbb{E}_{12}
  =
  \begin{pmatrix}
    0\\ 0\\ 1\\ 0
  \end{pmatrix},\hskip1ex
  \mathbb{E}_{21}
  =
  \begin{pmatrix}
    0\\ 0\\ 0\\ 1
  \end{pmatrix}.\n
\end{eqnarray}
The exciton COM degree of freedom can be solved as the problem of a particle in a periodic
potential. Therefore, the COM-coordinate basis function is given by the plane-wave
function
\begin{eqnarray}
  \chi_{\mathbf{G}}(\mathbf{R})
  =
  e^{-\mathtt{i}\mathbf{G}\cdot\mathbf{R}}
\end{eqnarray}
where $\mathbf{G}$ is reciprocal lattice vector. The exciton internal degree of freedom
can be treated as the eigenspectrum problem of an isolated exciton. It is known that STOs
can be used as the basis function to solve exciton internal wavefunctions accurately in
two dimensions\cite{zhang2019two, wu2019exciton, mypaper0, mypaper1}. The
internal-coordinate basis function is given by STOs, which have the form
\begin{eqnarray}
  \phi_{l_{\text{e}}l_{\text{h}},a}(\mathbf{r})
  =
  \left({e^{\mathtt{i}L_a\varphi}} / {\sqrt{2\pi}}\right)r^{N_a-1}
  \exp\left({-\mathcal{Z}_{l_{\text{e}}l_{\text{h}},a}r}\right)
\end{eqnarray}
with $N_a$ the shell number, $L_a$ the angular momentum, and
$\mathcal{Z}_{l_{\text{e}}l_{\text{h}},a}$ the shielding constant. Note that the phase
factor
$\exp\left(\mathtt{i}\bsym{\kappa}_{l_\text{e}l_\text{h}}\cdot\mathbf{r}\right)$ has
been added in Eq.~(\ref{exciton_wavefn}). This phase factor is utilized to shift the
internal momentum to center the origin point,
\begin{eqnarray}
  \exp\left(-{\mathtt{i}\bsym{\kappa}_{l_\text{e}l_\text{h}}\cdot\mathbf{r}}\right)
  \left(\mathbf{p}-\bsym{\kappa}_{l_\text{e}l_\text{h}}\right)
  \exp\left({\mathtt{i}\bsym{\kappa}_{l_\text{e}l_\text{h}}\cdot\mathbf{r}}\right)
  =\mathbf{p},
\end{eqnarray}
such that the matrix element can be written in a simpler form. Matrix elements of the
exciton Hamiltonian matrix and the overlap matrix span by the present basis are shown in
Appendix~\ref{sec:matrix_element}.

By using the exciton wavefunction, the radius and angular momentum of the hybridized
moir\'e exciton can be found. The exciton radius operator is defined by the matrix
representation
\begin{eqnarray}
  r_{\text{X}}(\mathbf{R},\mathbf{r})
  \equiv
  r
  \mathbb{I},
\end{eqnarray}
with
\begin{eqnarray}
  \mathbb{I}
  =
  \begin{pmatrix} 
    {1} & 0 & 0 & 0\\ 
    0 & {1} & 0 & 0\\ 
    0 & 0 & {1} & 0\\ 
    0 & 0 & 0 & {1} 
  \end{pmatrix}
\end{eqnarray}
a four-dimensional identity matrix. The exciton radius can be solved from
\begin{eqnarray}
  a_{\text{X},I}
  =
  \int
  \Psi^*_{\text{X},I\mathbf{0}}(\mathbf{R},\mathbf{r}) r_{\text{X}}(\mathbf{R},\mathbf{r})
  \Psi_{\text{X},I\mathbf{0}}(\mathbf{R},\mathbf{r})\text{d}^2R\text{d}^2r.\hskip2ex
\end{eqnarray}
The exciton angular momentum operator is defined as
$\mathcal{L}_{\text{X}}(\mathbf{R},\mathbf{r})
\equiv\mathcal{L}'_{\text{X}}(\mathbf{R})+\mathcal{L}''_{\text{X}}(\mathbf{r})$, where
\begin{eqnarray}
  \mathcal{L}'_{\text{X}}(\mathbf{R})
  =
  \left[\mathbf{R}\times(\mathbf{P}-\bsym{\mathcal{K}}_{l_{\text{e}}l_{\text{h}}})\right]
  \mathbb{I}
\end{eqnarray}
and
\begin{eqnarray}
  \mathcal{L}''_{\text{X}}(\mathbf{r})
  =
  \left[\mathbf{r}\times(\mathbf{p}-\bsym{\kappa}_{l_\text{e}l_\text{h}})\right]
  \mathbb{I}
\end{eqnarray}
are the matrix representations of the exciton COM angular momentum operator and the
exciton internal angular momentum operator. Since the internal angular momentum of
optically-active excitons should be zero, the root-mean-square exciton angular momentum is
used to measure the contribution of non-zero-angular-momentum Rydberg excitons on moir\'e
exciton wavefunction. The root-mean-square exciton angular momentum
$\sqrt{\braket{\mathcal{L}''^2_{\text{X}}}}$ is given by the formula
\begin{eqnarray}
  \braket{\mathcal{L}''^2_{\text{X}}}_{I}
  =
  \int
  \Psi^*_{\text{X},I\mathbf{0}}(\mathbf{R},\mathbf{r})  
  \mathcal{L}''^2_{\text{X}}(\mathbf{r})
  \Psi_{\text{X},I\mathbf{0}}(\mathbf{R},\mathbf{r})\text{d}^2R\text{d}^2r.\hskip2ex
\end{eqnarray}
Therefore, if the exciton wavefunction contains the STO
$\phi_{l_{\text{e}}l_{\text{h}},a}(\mathbf{r})$ with $L_a\neq{0}$, a non-zero value of
root-mean-square exciton angular momentum can be found.

\subsection{Optical absorption\label{ssec:optical_absorption}}

Based on Fermi's golden rule, the optical absorption spectrum can be given by the formula
\begin{eqnarray}
  \Gamma(\omega)
  &=&
  \frac{2\pi}{\omega}\sum_{I,\mu}|j^{\mu}_{0I}|^2\delta(\omega-E_{I}+E_{0}),
\end{eqnarray}
where ${E}_{I}$ the $I$-th state eigenenergy, $\bsym{j}_{II'} =
\braket{I|\hat{\bsym{j}}|I'}$ is the transition amplitude, with $\ket{I}$ the $I$-th state
ket. The transition between a one-exciton excited state and the ground state is known as
the one-exciton transition. By assuming the one-exciton excited states being excited
states $\ket{I}=\ket{\text{X}_{I\mathbf{K}}}$, the one-exciton transition amplitudes is
given by
\begin{eqnarray}
  \bsym{j}_{0,I\mathbf{K}}
  =
  \int\bsym{\mathcal{J}}_{\text{X}}(\mathbf{r})
  \Psi_{\text{X},I\mathbf{K}}(\mathbf{R},\mathbf{r})\text{d}^2r\text{d}^2R,
  \label{transition_amplitude}
\end{eqnarray}
where $\bsym{\mathcal{J}}_{\text{X}}(\mathbf{r}) = \begin{pmatrix}
\bsym{\mathcal{J}}_{11}(\mathbf{r}) & \bsym{\mathcal{J}}_{22}(\mathbf{r}) & 0 & 0
\end{pmatrix}$, with $\bsym{\mathcal{J}}_{ll}(\mathbf{r})$ the momentum matrix element.
The momentum matrix element is given by
\begin{eqnarray}
  {\mathcal{J}}^{\pm}_{ll}(\mathbf{r})
  &=&
  e\big[{\mathcal{J}}^{x}_{ll}(\mathbf{r})
  \pm\mathtt{i}{\mathcal{J}}^{y}_{ll}(\mathbf{r})\big]/\sqrt{2}
  =
  \delta_{\tau,\pm}\delta(\mathbf{r})ev_{\text{F},l},
\end{eqnarray}
where $v_{\text{F},l}$ is the Fermi velocity for each layer $l$. Note that $\tau$ is the
valley index and the matrix element is circular-polarization selected. However, this
selection can be ignored since we have not included the valley degree of freedom in this
model. By using the exciton wavefunction, the one-exciton transition amplitude is given by
\begin{eqnarray}
  j^{\pm}_{0,I\mathbf{K}}
  &=&
  2\pi\sum_{l,a}\sum_{\mathbf{G}}ev_{\text{F},l}\delta_{\mathbf{K},\mathbf{G}}
  C_{(ll,a,\mathbf{G}),I\mathbf{G}}\phi_{ll,a}(\mathbf{0}).
  \label{transition_amplitude}
\end{eqnarray}
Since only the first Brillouin zone is considered in the calculation, the selection for
the COM wavevector is given by $\mathbf{K}=\mathbf{0}$ for optical absorption. The
ground-state energy can be assigned as $E_{0}=0$ and the excited-state energy for every
one-exciton excited state is given by $E_{I}=\varepsilon_{\text{X},I\mathbf{0}}$.
Combining all the information, the optical absorption spectrum can be rewritten as
\begin{eqnarray}
  \Gamma(\omega)
  &=&
  \sum_{I,\tau=\pm}\frac{|j^{\tau}_{0,I\mathbf{0}}|^2}{\varepsilon_{\text{X},I\mathbf{0}}}
  \frac{2\eta}{(\omega-\varepsilon_{\text{X},I\mathbf{0}})^2+\eta^2},
  \label{spectral_density}
\end{eqnarray}
where $\eta$ is a line-broadening factor. Eq.~(\ref{spectral_density}) is the working
formula for us to calculate the optical absorption spectrum of moir\'e heterobilayers.
Note that if $C_{(ll,a,\mathbf{G}),I\mathbf{G}} =\delta_{l,1}
C_{(11,a,\mathbf{G}),I\mathbf{G}}$ in Eq.~(\ref{transition_amplitude}), the formula in
Eq.~(\ref{spectral_density}) can be reduced to the optical absorption formula for
intralayer exciton transitions in the layer $l=1$. The interlayer exciton transition is
only introduced through the wavefunction coefficient $C_{(ll,a,\mathbf{G}),I\mathbf{G}}$
of the $I$-th exciton state, where each exciton state is a linear combination of
intralayer and interlayer exciton wavefunctions.

\section{Applications\label{sec:application}}

\begin{table*}
\centering
\begin{tabular}{c | c c c c c c c c c | c c}
\hline\hline
Materials & $E_{\text{gap}}$(eV)
& $E^{\text{v}}_{\text{edge}}$(eV) & $E^{\text{c}}_{\text{edge}}$(eV)
& $E^{\text{v}}_{\text{SO}}$(meV) & $E^{\text{c}}_{\text{SO}}$(meV)
& $a$(\AA) & $m_{\text{e}}/m_0$ & $m_{\text{h}}/m_0$
& $r_0$(\AA)
& $E_{\text{X}}$(meV) & ${10}^{3}v_{\text{F}}/c_0$\\
\hline
{$\text{MoSe}_2$} & {$1.874$} & {$-5.750$}& $-3.876$ & {$-184$} & {$20$} & {$3.288$}
& {$0.70$} & {$0.70$} & {$39$} & $232$ & {$1.62$} \\
{$\text{WS}_2$} & {$2.238$} & {$-6.190$} & $-3.952$ & {$-425$} & {$-31$} & {$3.154$}
& {$0.35$} & {$0.35$} & {$34$} & $177$ & {$2.50$} \\
{$\text{WSe}_2$} & {$1.890$} & {$-5.490$} & $-3.600$ & {$-462$} & {$-37$} & {$3.286$}
& {$0.40$} & {$0.40$} & {$45$} & $165$ & {$2.15$} \\
\hline\hline
\end{tabular}
\caption{Effective-mass-model parameters for TMDC monolayers. The dielectric constants for
the modified Rytova-Keldysh potential are given by $\kappa_{1}=4.4$ and $\kappa_0=2.0$.
$E_{\text{gap}}$ is the band-gap energy, $E^{\text{v}}_{\text{edge}}$ is the valence
band-edge energy, $E^{\text{c}}_{\text{edge}}$ is the conduction band-edge energy,
$E^{\text{v}}_{\text{SO}}$ is the valence band spin-orbit splitting energy,
$E^{\text{c}}_{\text{SO}}$ is the conduction band spin-orbit splitting energy, $a$ is the
lattice constant of the TMDC monolayer, $m_{\text{e}}$ and $m_{\text{h}}$ are effective
electron and hole masses, $m_0$ is the bare electron mass, $r_0$ is the screening length,
$E_{\text{X}}$ is the exciton binding energy for the $1s$-orbital exciton, $v_{\text{F}}$
is the Fermi velocity, and $c_0$ is the light velocity. The exciton binding energy is
calculated by variationally solving the effective mass model with Rytova-Keldysh potential
as the screened Coulomb potential. The Fermi velocity is estimated by
$E_{\text{gap}}=(m_{\text{e}}+m_{\text{h}})v_{\text{F}}$. The effective masses and
screening lengths are acquired from Ref.~\onlinecite{goryca2019revealing}. The lattice
constants are acquired from Ref.~\onlinecite{wilson1969transition}. The band-edge energies
are acquired from Ref.~\onlinecite{zhang2016systematic}. The spin-orbit splitting energies
are acquired from Ref.~\onlinecite{kormanyos2015k}.}
\label{tab:TMDCs}
\end{table*}

In this section, the exciton continuum model is applied to the simulation of optical
absorption spectra of $\text{WSe}_2$/$\text{WS}_2$ heterobilayers and
$\text{MoSe}_2$/$\text{WS}_2$ heterobilayers. Twist-angle and electric-field dependences
of optical absorption spectra are calculated and discussed. Parameters for the
effective-mass band and the modified Rytova-Keldysh potential of heterobilayers embedded
in hexagonal boron nitride substrates are given in Table~\ref{tab:TMDCs}. All other
parameters, including moir\'e potential depths and charge-transfer coupling strengths, are
chosen to fit simulated optical spectra with experimental observations.

\subsection{Excitons in $\text{WSe}_2$/$\text{WS}_2$ heterobilayers\label{ssec:WSe2_WS2}}

\begin{figure}
\includegraphics[width=0.95\linewidth]{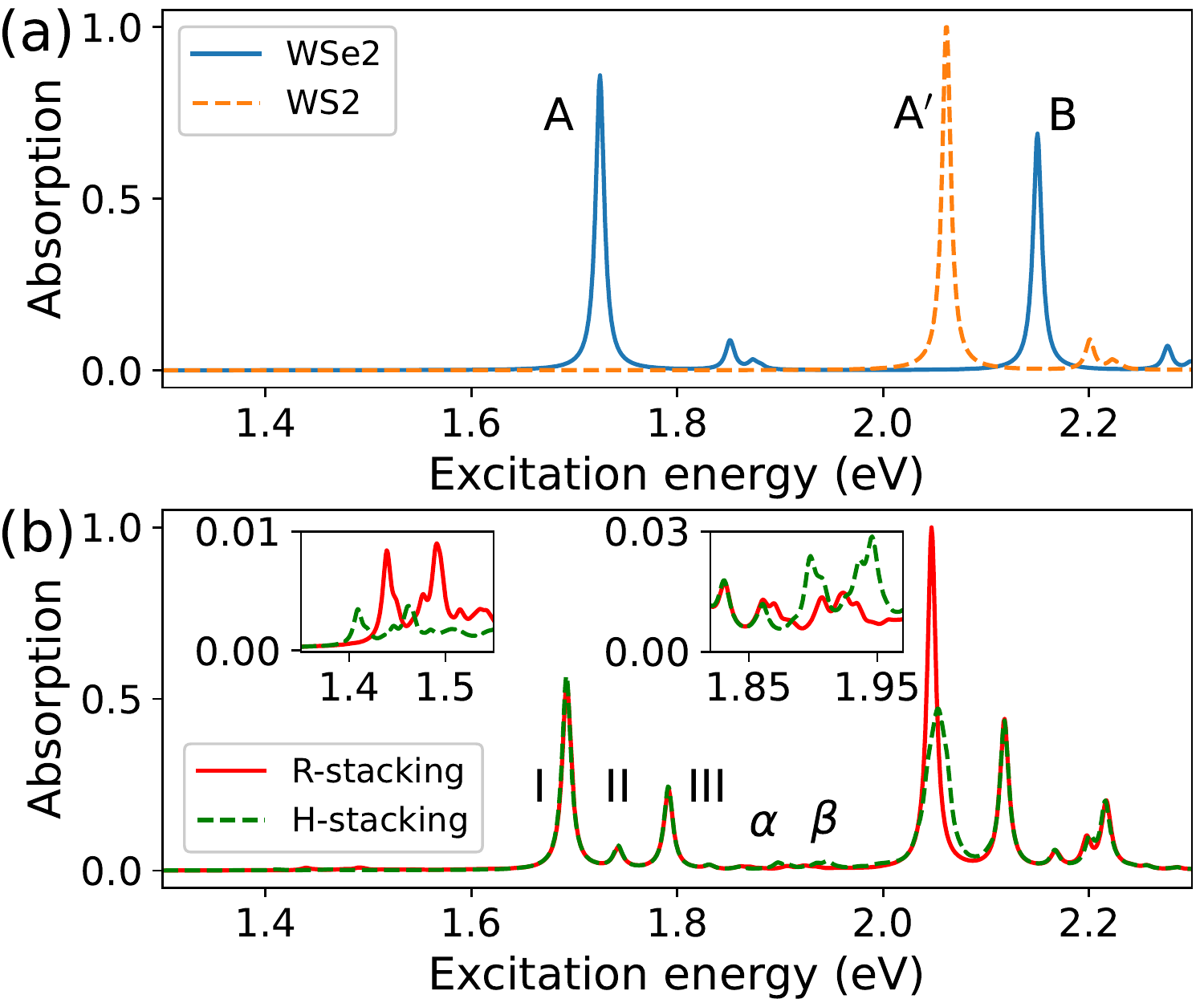}
\caption{(a) Calculated absorption spectra of $\text{WSe}_2$ and $\text{WS}_2$ monolayers.
(b) Calculated absorption spectra of R-stacked and H-stacked $\text{WSe}_2$/$\text{WS}_2$
heterobilayers, with parameters $w_{\text{e}}=w_{\text{h}}=20$ meV, $V_{\text{WSe}_2}=30$
meV, $V_{\text{WS}_2}=5$ meV, $\eta=5$ meV, and $d=7.0$ \AA.}
\label{fig:spectrum_WSe2_WS2}
\end{figure}

\begin{table}
\centering
\begin{tabular}{c c c c c c}
\hline\hline
Excitons & $E_{exc}$ (eV) & $E_{\text{X}}$ (meV) & $|j_{\text{X}}|^2$ (a.u.)
& $a_{\text{X}}$ (\AA) & $\sqrt{\braket{\mathcal{L}''^2_{\text{X}}}}$\\
\hline
$\text{A-}{1s}$ & $1.725$ & $165$ & $1.000$ & $14.0$ & $0$\\
$\text{A-}{2s}$ & $1.851$ & $39$  & $0.107$ & $66.8$ & $0$\\
$\text{A-}{2p}$ & $1.830$ & $50$  & $0$     & $43.0$ & $1$\\
\hline
I        & $1.690$ & $195$ & $0.763$ & $19.9$ & $0.395$ \\
II       & $1.739$ & $144$ & $0.088$ & $16.5$ & $0.261$ \\
III      & $1.789$ & $98$  & $0.337$ & $39.0$ & $0.603$ \\
$\alpha$ & $1.898$ & $119$ & $0.025$ & $38.0$ & $0.660$ \\
$\beta$  & $1.947$ & $69$  & $0.017$ & $56.9$ & $1.001$ \\
\hline\hline
\end{tabular}
\caption{Calculated properties of A excitons in $\text{WSe}_2$ monolayer (top three rows)
and moir\'e excitons in H-stacked $\text{WSe}_2$/$\text{WS}_2$ heterobilayers (remaining
rows). $E_{exc}$ is the exciton transition energy, $E_{\text{X}}$ is the exciton binding
energy, $|j_{\text{X}}|^2$ is the transition amplitude, $a_{\text{X}}$ is the exciton
radius, and $\sqrt{\braket{\mathcal{L}''^2_{\text{X}}}}$ is the exciton root-mean-square
angular momentum. The properties of intralayer moir\'e excitons are calculated without
considering the interlayer tunneling.}
\label{tab:excitons}
\end{table}

\begin{figure}
\includegraphics[width=0.95\linewidth]{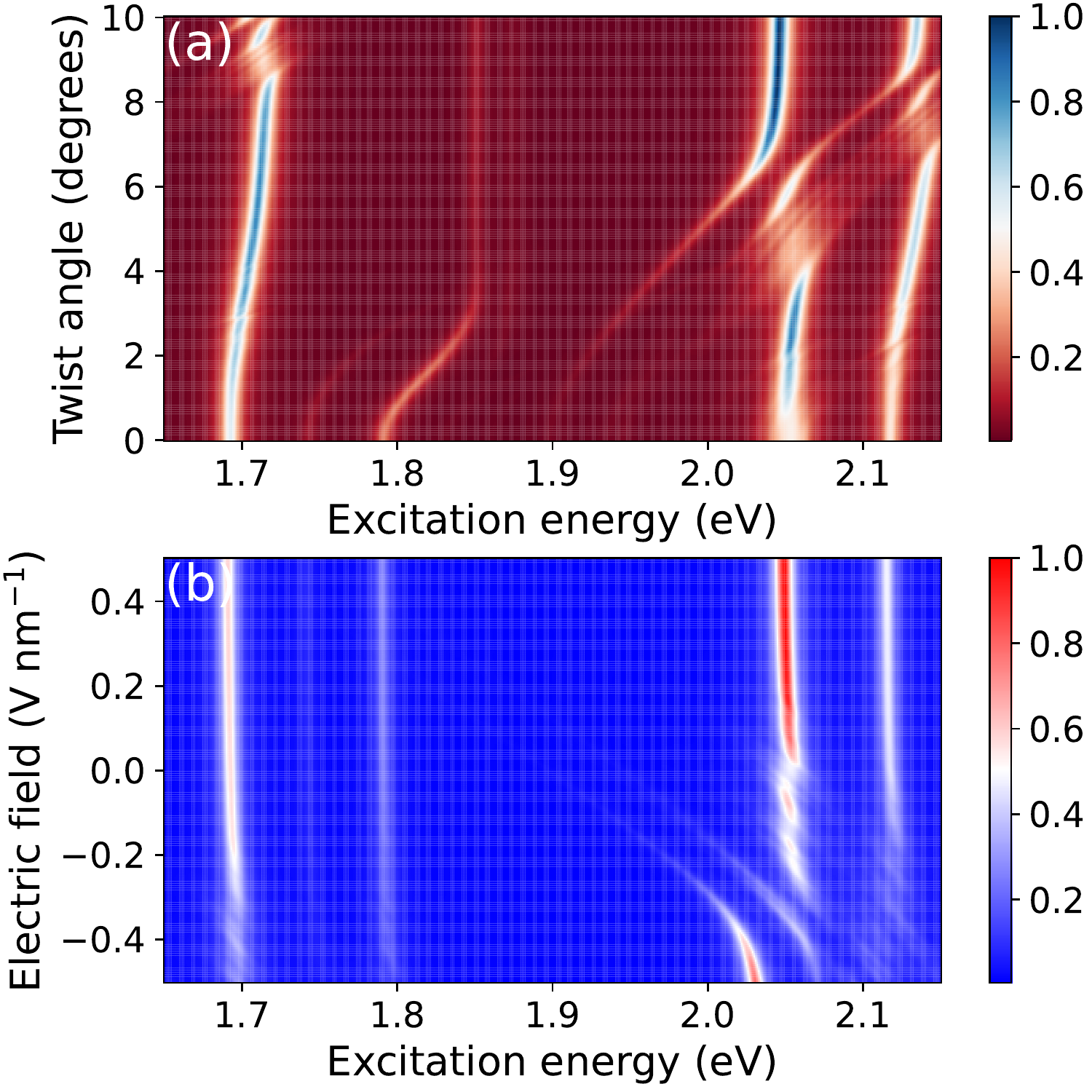}
\caption{(a) Calculated twist-angle dependence and (b) electric-field dependence of
optical absorption spectra of H-stacked $\text{WSe}_2$/$\text{WS}_2$ heterobilayers.}
\label{fig:twist_electric_WSe2_WS2}
\end{figure}

Calculated absorption spectra of $\text{WSe}_2$ monolayer, $\text{WS}_2$ monolayer, and
$\text{WSe}_2$/$\text{WS}_2$ heterobilayers are shown in Fig.~\ref{fig:spectrum_WSe2_WS2}
(a) and (b). The contributions from A and B excitons in $\text{WSe}_2$ monolayer and A
excitons in $\text{WS}_2$ monolayer have been included in Fig.~\ref{fig:spectrum_WSe2_WS2}
(a). Signatures of $2s$, $3s$ Rydberg excitons in $\text{WSe}_2$ monolayer are also shown
around $1.85$ eV for A excitons and around $2.25$ eV for B excitons as the satellite peaks
of the dominated $1s$ exciton transition peak. By including moir\'e potentials and
charge-transfer couplings in the exciton calculation for $\text{WSe}_2$/$\text{WS}_2$
heterobilayers, signatures of intralayer and interlayer moir\'e excitons are shown. The
three prominent peaks around $1.7$-$1.8$ eV, labeled I, II, and III peaks as in
Ref.~\onlinecite{jin2019observation}, can be attributed to the intralayer moir\'e excitons
extending from A excitons of $\text{WSe}_2$. The inset figures in
Fig.~\ref{fig:spectrum_WSe2_WS2} (b) show enlarged peaks from optical absorption by
interlayer excitons. The signature around $1.4$-$1.5$ eV in the spectrum of the R-stacked
heterobilayer can be attributed to the interlayer moir\'e excitons with the electron
locating at $\mathbf{K}$ valley of $\text{WS}_2$ monolayer and the hole locating at
$\mathbf{K}$ valley of $\text{WSe}_2$ monolayer. The signature around $1.4$-$1.5$ eV in
the spectrum of the H-stacked heterobilayer can be attributed to the interlayer moir\'e
excitons with the electron locating at $\mathbf{K}'$ valley of $\text{WS}_2$ monolayer and
the hole locating at $\mathbf{K}$ valley of $\text{WSe}_2$ monolayer. The signature around
$1.85$-$1.95$ eV in the spectrum of the H-stacked heterobilayer can be attributed to the
interlayer moir\'e excitons with the electron locating at $\mathbf{K}$ valley of
$\text{WS}_2$ monolayer and the hole locating at $\mathbf{K}'$ valley of $\text{WSe}_2$
monolayer. Interlayer-exciton transition peaks are much less obvious in comparison with
intralayer-exciton transition peaks because of the large energy difference between
interlayer excitons and intralayer excitons. These signatures are overall coincident with
experimental observations in the literature\cite{jin2019identification, tang2021tuning}.

In Table~\ref{tab:excitons}, exciton transition energies, exciton binding energies,
transition amplitudes, exciton radii, and exciton root-mean-square angular momentums of
exciton transitions in $\text{WSe}_2$ monolayer and H-stacked $\text{WSe}_2$/$\text{WS}_2$
heterobilayers are shown. Excitons $\alpha$, $\beta$ are two interlayer-exciton
transitions of H-stacked $\text{WSe}_2$/$\text{WS}_2$ heterobilayers, as indicated in
Fig.~\ref{fig:spectrum_WSe2_WS2} (b). Because the internal degree of freedom is included
in the exciton calculation, intralayer excitons and interlayer excitons can have different
radii and angular momentum. According to the variational exciton wavefunction in
Eq.~(\ref{exciton_wavefn}), an exciton in heterobilayers can be seen as the hybridization
of various Rydberg excitons in monolayers. Therefore, hybridized excitons listed in
Table~\ref{tab:excitons} show mixture properties from different Rydberg excitons in
$\text{WSe}_2$ monolayer.

In Fig.~\ref{fig:twist_electric_WSe2_WS2} (a), (b), calculated twist-angle and
electric-field dependences of optical absorption spectra in H-stacked
$\text{WSe}_2$/$\text{WS}_2$ heterobilayers are shown. For the twist-angle dependence, the
three intralayer-exciton peaks I, II, and III show spectral shifts with increasing twist
angles. The peak I converges to the A-$1s$ exciton of $\text{WSe}_2$ monolayer, the peak
III converges to the A-$2s$ exciton of $\text{WSe}_2$ monolayer, and the peak II
diminishes. For the electric-field dependence, an out-of-plane electric field is included
in the model by modulating the band-edge energy of $\text{WS}_2$ monolayer as
\begin{eqnarray}
  \tilde{\epsilon}_{\text{e},\text{WS}_2}
  =
  \epsilon_{\text{e},\text{WS}_2}-\xi_{z}F_{z},\hskip2ex
  \tilde{\epsilon}_{\text{h},\text{WS}_2}
  =
  \epsilon_{\text{h},\text{WS}_2}-\xi_{z}F_{z},\hskip2ex
  \label{electric_field}
\end{eqnarray}
where $\epsilon_{\text{e},\text{WS}_2}$ and $\epsilon_{\text{h},\text{WS}_2}$ are
band-edge energies for the conduction band and the valence band, $\xi_{z}$ is the
out-of-plane electric dipole moment, and $F_{z}$ is the out-of-plane electric field
strength. The electric dipole moment is assigned as $\xi_{z}=0.4\;e\cdot$nm, where $e$
denotes the elementary charge. Signatures of hybridization can be found in both the
twist-angle dependence and the electric-field dependence, particularly near the A-exciton
transition energy of $\text{WS}_2$ monolayer. As shown in
Fig.~\ref{fig:twist_electric_WSe2_WS2} (a), signatures of avoiding crossing are found at
$7^\circ$ twist angle and around $2.1$ eV excitation energy. In
Fig.~\ref{fig:twist_electric_WSe2_WS2} (b), signatures of avoiding crossing are found
around $-0.3\sim-0.5$ V nm${}^{-1}$ electric field and around $2.1$ eV excitation energy.
These signatures indicate the hybridization between intralayer excitons and interlayer
excitons through charge-transfer couplings as the excitation energies of intralayer
excitons and interlayer excitons are close.

Note that, for real-world experiments, the moir\'e heterobilayer is embedded in a
dielectric substrate, which is normally composed of multiple layers of hexagonal boron
nitrides (hBNs) or silicon dioxide (SiO${}_2$), and the applied external electric field
can be screened by the substrate. Therefore, the out-of-plane electric field mentioned
above is the out-of-plane displacement field and the simulated electric-field dependence
might be different from the experimental observation due to different dielectric screening
for different experimental conditions.

\subsection{Excitons in $\text{MoSe}_2$/$\text{WS}_2$ heterobilayers\label{ssec:MoSe2_WS2}}

\begin{figure}
\includegraphics[width=0.95\linewidth]{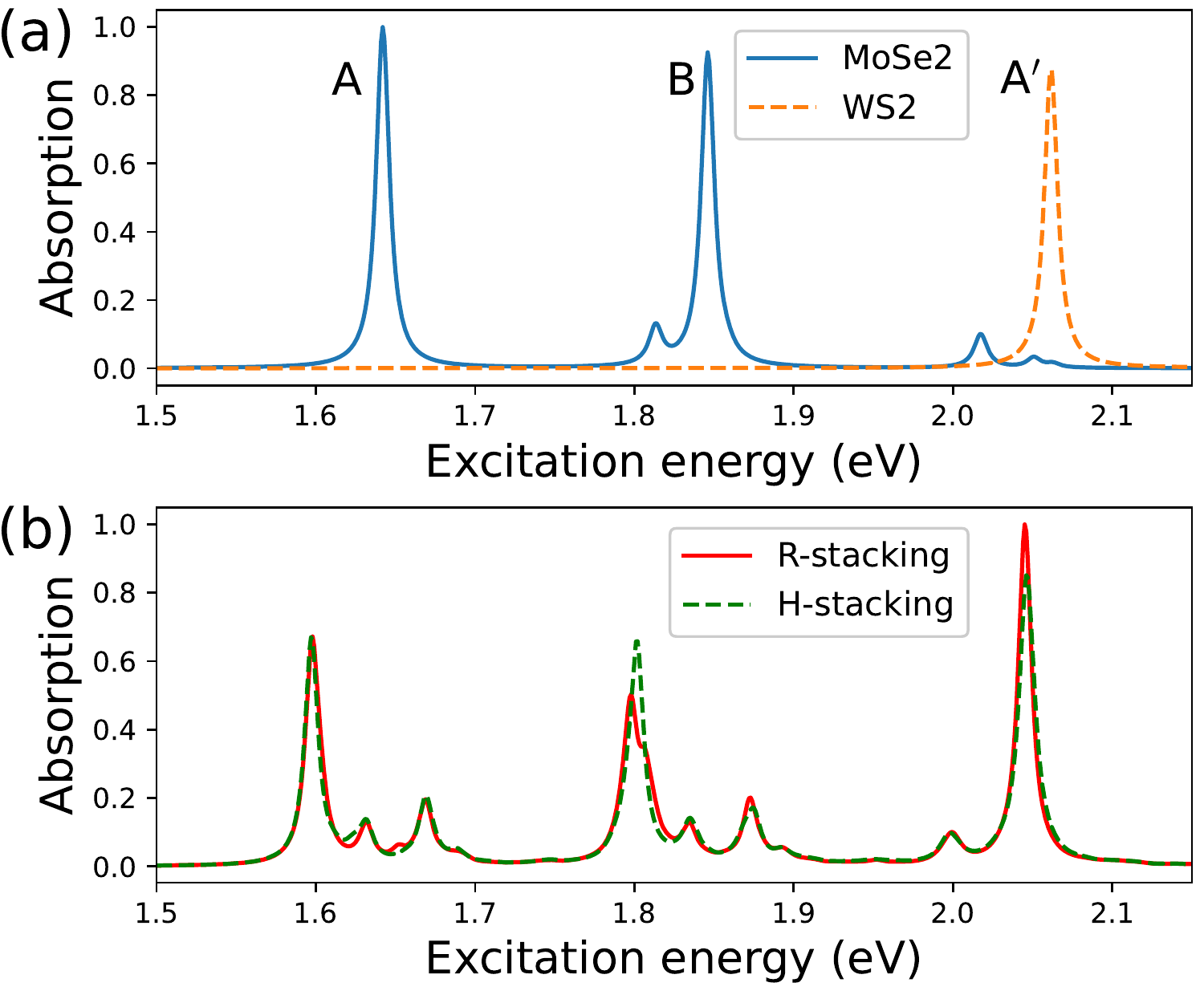}
\caption{(a) Calculated absorption spectra of $\text{MoSe}_2$ and $\text{WS}_2$
monolayers. (b) Calculated absorption spectra of R-stacked and H-stacked
$\text{MoSe}_2$/$\text{WS}_2$ heterobilayers, with parameters
$w_{\text{e}}=w_{\text{h}}=10$ meV, $V_{\text{MoSe}_2}=35$ meV, $V_{\text{WS}_2}=5$ meV,
$\eta=5$ meV, and $d=7.0$ \AA.}
\label{fig:spectrum_MoSe2_WS2}
\end{figure}

\begin{figure}
\includegraphics[width=0.95\linewidth]{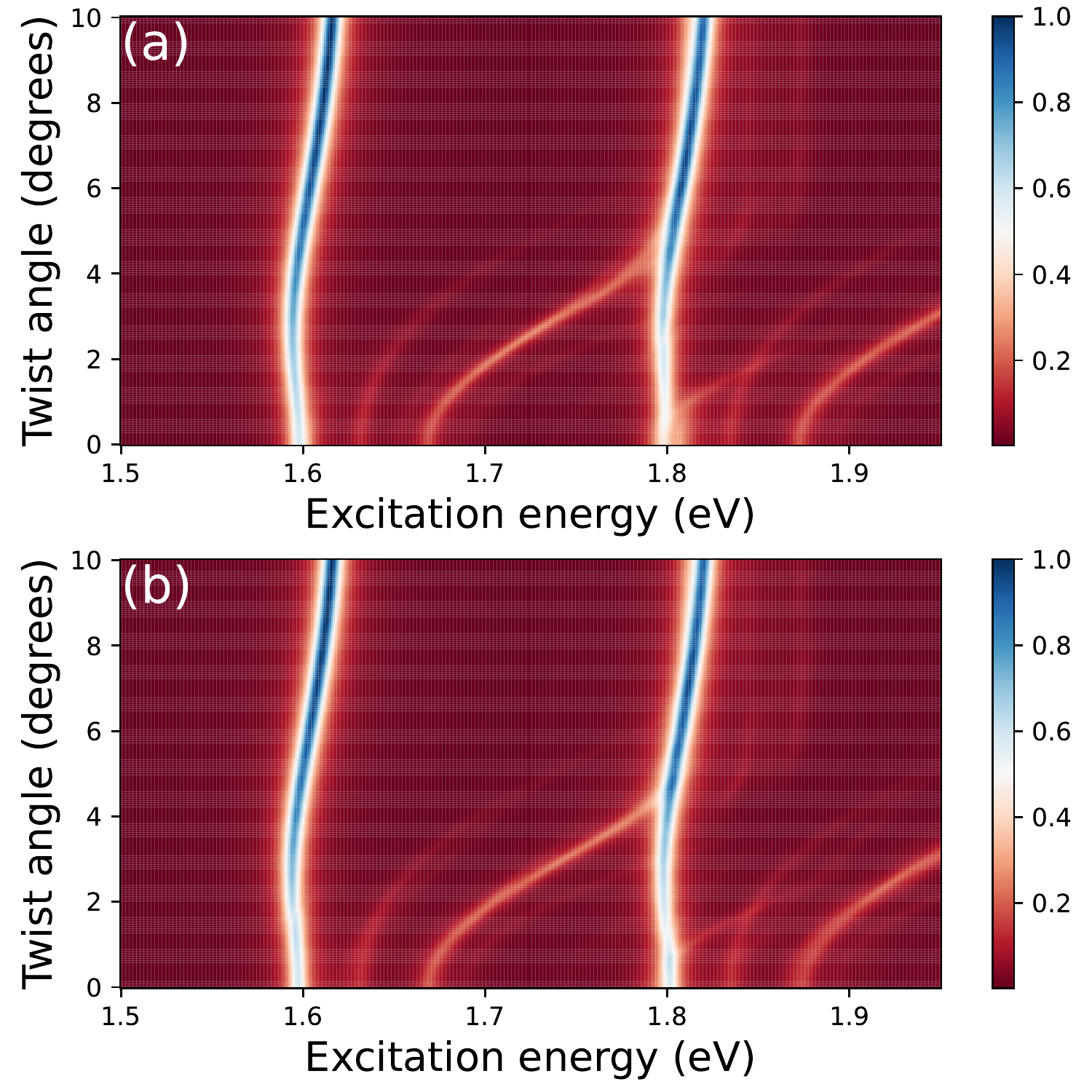}
\caption{Calculated twist-angle-dependent absorption spectra of (a) R-stacked
$\text{MoSe}_2$/$\text{WS}_2$ heterobilayers and (b) H-stacked
$\text{MoSe}_2$/$\text{WS}_2$ heterobilayers.}
\label{fig:twist_angle_MoSe2_WS2}
\end{figure}

\begin{figure}
\includegraphics[width=0.95\linewidth]{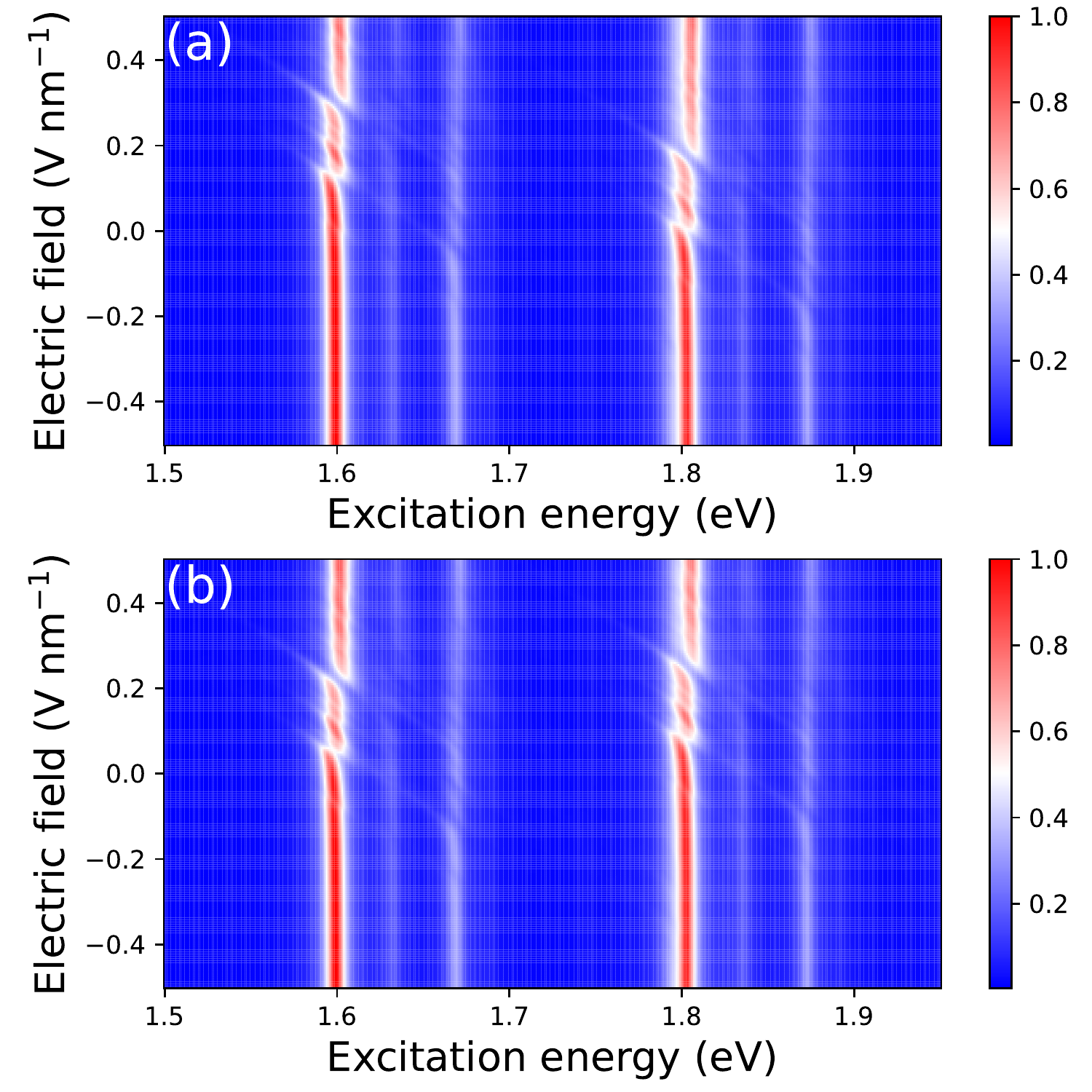}
\caption{Calculated electric-field-dependent absorption spectra of (a) R-stacked
$\text{MoSe}_2$/$\text{WS}_2$ heterobilayers and (b) H-stacked
$\text{MoSe}_2$/$\text{WS}_2$ heterobilayers.}
\label{fig:electric_field_MoSe2_WS2}
\end{figure}

Calculated absorption spectra of $\text{MoSe}_2$ monolayer, $\text{WS}_2$ monolayer,
R-stacked and H-stacked $\text{MoSe}_2$/$\text{WS}_2$ heterobilayers are shown in
Fig.~\ref{fig:spectrum_MoSe2_WS2} (a) and (b). Again, optical absorption signatures of A
excitons, B excitons, and $2s$, $3s$ Rydberg excitons of $\text{MoSe}_2$ monolayer and
$\text{WS}_2$ monolayer are shown in Fig.~\ref{fig:spectrum_MoSe2_WS2} (a). The
interlayer-exciton transition energy of the R-stacked heterobilayer is close to the
A-exciton transition energy of $\text{MoSe}_2$ monolayer and the interlayer-exciton
transition energy of the H-stacked heterobilayer is close to the B-exciton transition
energy of $\text{MoSe}_2$ monolayer. The hybridization between the intralayer excitons and
the interlayer excitons induces small changes in the shape of peaks around the A-exciton
transition energy of $\text{MoSe}_2$ monolayer ($1.6$-$1.7$ eV) for the R-stacked
heterobilayer and around the B-exciton transition energy of $\text{MoSe}_2$ monolayer
($1.8$-$1.9$ eV) for the H-stacked heterobilayer.

In Fig.~\ref{fig:twist_angle_MoSe2_WS2}, calculated twist-angle-dependent absorption
spectra of R-stacked $\text{MoSe}_2$/$\text{WS}_2$ heterobilayers and H-stacked
$\text{MoSe}_2$/$\text{WS}_2$ heterobilayers are shown, and in
Fig.~\ref{fig:electric_field_MoSe2_WS2}, calculated electric-field-dependent absorption
spectra of R-stacked $\text{MoSe}_2$/$\text{WS}_2$ heterobilayers and H-stacked
$\text{MoSe}_2$/$\text{WS}_2$ heterobilayers are shown. The out-of-plane electric field is
included in the model by the same band-edge modulation of $\text{WS}_2$ monolayer in
Eq.~(\ref{electric_field}). Signatures of avoiding crossing due to exciton hybridization
can be found in both simulated spectra. Experimental twist-angle-dependent absorption
spectra of R-stacked and H-stacked $\text{MoSe}_2$/$\text{WS}_2$ heterobilayers can be
found in Ref.~\onlinecite{alexeev2019resonantly}. Experimental electric-field-dependent
absorption spectra of R-stacked $\text{MoSe}_2$/$\text{WS}_2$ heterobilayers can be found
in Ref.~\onlinecite{tang2021tuning}. The simulation and the observation share resemblances
but also noticeable differences. Those differences might be attributed to the simplicity
of the present model. Particularly, intervalley exciton exchange, spin-flip interlayer
charge transfer, and electron-phonon coupling are ignored. Some exciton transition
signatures could be lost due to the omission. However, since it is important to test the
limit of the proposed model and the context of this article is already quite ample, we
will leave those topics to future works.

\section{Discussions and conclusion\label{sec:discussion}}

A critical issue we have not discussed is the properness of parameters chosen in the
exciton continuum model. Apart from the parameters also used in monolayers, four
parameters decide the optical spectra of heterobilayers: two moir\'e potential depths
$V_{1}$, $V_{2}$ and two charge-transfer coupling strength $w_{\text{e}}$, $w_{\text{h}}$.
These parameters can be obtained by fitting observed optical spectra as what we have done
in this work, or calculated by atomistic simulation. However, there is an inconsistency
between the moir\'e potential depths acquired from atomistic simulation and fitting the
spectrum. By DFT calculation, moir\'e potential depths on $\text{MoSe}_2$ layer and
$\text{WSe}_2$ layer of different heterobilayers are estimated to be around ${8}\sim{10}$
meV (peak-to-peak energy difference around $80\sim{100}$ meV)\cite{naik2022intralayer}.
Via various simulations including this work by exciton continuum models, observed
moir\'e-exciton signatures can only be explained by larger moir\'e potential depths about
$15\sim{35}$ meV\cite{jin2019observation, lin2023remarkably}. The discrepancy may be
attributed to the oversimplification of the effective moir\'e potential, the insufficiency
of DFT calculation to simulate atom-atom potential without sufficient good van der Waals
density functionals\cite{rydberg2003van}, or the underestimation of geometry relaxation
and strain in moir\'e heterobilayer\cite{lin2023remarkably}. A theoretical work eliminates
the oversimplification by using DFT calculation to simulate moir\'e excitons
atomistically\cite{naik2022intralayer}. They found that one of the moir\'e-exciton
signatures can be attributed to the charge-transfer exciton across different moir\'e unit
cells, and the low moir\'e potential depth ($\sim{9}$ meV) is sufficient to explain all
intralayer moir\'e-exciton signatures. However, we do not reach the same conclusion since
all radii of intralayer moir\'e excitons calculated in this work and listed in
Table~\ref{tab:excitons} are shorter than the moir\'e lattice constant ($a_{\text{M}}\sim
80$ \AA) of $\text{WSe}_2$/$\text{WS}_2$ heterobilayers. A scanning tunneling microscopy
experiment on twisted TMDC heterobilayers also indicates that DFT calculations might
underestimate the moir\'e potential depths\cite{shabani2021deep}. Further investigation on
this issue is required in future works.

Note that the exciton continuum model in this study is only applied to the simulation of
the optical spectrum of $\text{WSe}_2$/$\text{WS}_2$ and $\text{MoSe}_2$/$\text{WS}_2$
heterobilayers with small twist angles. This is because the present model is still too
simple to accurately simulate twisted TMDC homobilayers or twisted TMDC heterobilayers
with large twist angles. Several critical elements are not considered in this model. One
is the valley degree of freedom of TMDCs, which is related to the optical selection rule
of circular-polarized light. Another is the mirror-symmetry-breaking interaction, which
should exist in chiral or helical materials. Without these elements, the chirality and
optical activity of twisted TMDC bilayers can not be described correctly, and thus the
electronic structure of twisted TMDC bilayers may not be simulated accurately by this
model. Additionally, as mentioned in Sec.~\ref{ssec:MoSe2_WS2}, intervalley exciton
exchange and spin-flip interlayer charge transfer are ignored in this model. These effects
are also related to the valley degree of freedom and this ignorance might also contribute
to inaccuracy. Since this article is already lengthy, we decided to keep the simplicity of
this model and omit these elements. In future works, we intend to include the valley
degree of freedom and the mirror-symmetry-breaking interaction in an improved model and
study a broader range of twisted TMDC bilayers.

Although this exciton continuum model might suffer from oversimplification, it can be
beneficial with its simplicity for future extended studies. For example, in this work,
only lattice mismatch and twist angle on the moir\'e pattern are considered, and the
rotational symmetry of the system is not broken. However, in reality, the threefold
rotational symmetry of moir\'e heterobilayers could be broken in the presence of the
uniaxial strain, and the optical properties of interlayer and hybrid excitons would be
affected by the symmetry breaking. The modification could be done by including a
pseudo-gauge potential in terms of strain tensor components\cite{zheng2021twist,
zheng2022anomalous}. Another example is that exciton complexes such as trions and
biexcitons are known to be computationally exhausting to simulate. It is easier to use
effective-mass models and variational wavefunctions to study exciton
complexes\cite{mypaper0, mypaper1}. It is a natural extension of our theory to build the
theoretical framework of using continuum models and variational wavefunctions to study
hybridized moir\'e exciton complexes in multilayer materials. Additionally, with the
future inclusion of intervalley exchange and electron-phonon coupling into this model,
charge and exciton dynamics in moir\'e materials could also be studied without too
exhausted computation.

In conclusion, we propose a continuum model to study hybridized moir\'e excitons in TMDC
heterobilayers. A variational wavefunction method is used to solve this model. We use the
present theory to study optical absorption spectra of $\text{WSe}_2$/$\text{WS}_2$ and
$\text{MoSe}_2$/$\text{WS}_2$ heterobilayers. A good correspondence between theoretical
simulation and experimental observation of moir\'e excitons is found. This model could set
the foundation of a theoretical framework to study the physical properties of moir\'e
excitons in different moir\'e materials.

\begin{acknowledgments}

This work was supported by the National Science and Technology Council, the Ministry of
Education (Higher Education Sprout Project NTU-112L104022), and the National Center for
Theoretical Sciences of Taiwan.

\end{acknowledgments}

\appendix

\section{Modified Rytova-Keldysh potential\label{sec:RK_potential}}

The modified Rytova-Keldysh potential for parallel bilayers is derived in this section.
The notation and formulation of the present derivation are taken from
Ref.~\onlinecite{schwinger1998classical}.

\subsection{Reduced Green's function}

The Poisson equation for electrostatics in three-dimensional (3D) free space is written as
\begin{eqnarray}
  \nabla^2\Phi(\mathbf{r})=-{\varrho(\mathbf{r})}/{\kappa},
\end{eqnarray}
where $\Phi(\mathbf{r})$ is the electrostatic potential, $\varrho(\mathbf{r})$ is the
charge density, and $\kappa$ is the dielectric constant. The charge-free electrostatic
potential can be solved by the Laplace equation, $\nabla^2\Phi(\mathbf{r})=0$. The formal
solution of the Poisson equation in three-dimensional free space is given by
\begin{eqnarray}
  \Phi(\mathbf{r})
  =
  \frac{1}{\kappa}\int\frac{\varrho(\mathbf{r}')}{|\mathbf{r}-\mathbf{r}'|}\text{d}^3r.
\end{eqnarray}
The Green's function at a point $\mathbf{r}$ interacting with the point charge $e$ at
$\mathbf{r}'$ in 3D space is given by
\begin{eqnarray}
  G(\mathbf{r},\mathbf{r}')
  =
  \frac{1}{\kappa|\mathbf{r}-\mathbf{r}'|}.
\end{eqnarray}
Since $\nabla^2(1/r)=-4\pi\delta(\mathbf{r})$ and thus
\begin{eqnarray}
  \kappa\nabla^2G(\mathbf{r},\mathbf{r}')=-4\pi\delta(\mathbf{r}-\mathbf{r}'),
  \label{green_fn_eq}
\end{eqnarray}
Green's function is a solution of the Poisson equation. The Green's function can also
be derived from solving Eq.~(\ref{green_fn_eq}). By using the property of the delta 
function,
\begin{eqnarray}
  \delta(\mathbf{r}-\mathbf{r}')
  =
  \int\frac{\text{d}^3k}{(2\pi)^3}e^{\mathtt{i}\mathbf{k}\cdot(\mathbf{r}-\mathbf{r}')},
\end{eqnarray}
and the derivative $\bsym{\nabla}e^{\mathtt{i}\mathbf{k}\cdot(\mathbf{r}-\mathbf{r}')} =
\mathtt{i}\mathbf{k}e^{\mathtt{i}\mathbf{k}\cdot(\mathbf{r}-\mathbf{r}')}$, the Fourier
transform of Green's function can be found as
\begin{eqnarray}
  G(\mathbf{r},\mathbf{r}')
  =
  \int\frac{\text{d}^3k}{(2\pi)^3}
  \frac{4\pi}{\kappa{k}^2}e^{\mathtt{i}\mathbf{k}\cdot(\mathbf{r}-\mathbf{r}')}.
  \label{green_fn_k}
\end{eqnarray}
Singling out the $z$ direction in a 3D free space, Eq.~(\ref{green_fn_k}) can be rewritten
as
\begin{eqnarray}
  G(\mathbf{r},\mathbf{r}')
  &=&
  \frac{4\pi}{\kappa}\int\frac{\text{d}k_{x}\text{d}k_{y}}{(2\pi)^2}
  e^{\mathtt{i}\left[k_{x}(x-x')+k_{y}(y-y')\right]}\n
  &&\times\int\frac{\text{d}k_{z}}{2\pi}\frac{e^{\mathtt{i}k_{z}(z-z')}}
  {k^2_{x}+k^2_{y}+k^2_{z}}.
\end{eqnarray}
By using
\begin{eqnarray}
  \int^{\infty}_{-\infty}\frac{\text{d}k_{z}}{2\pi}
  \frac{e^{\mathtt{i}k_{z}(z-z')}}{{k}_{\parallel}^2+k^2_{z}}
  =
  \frac{1}{2{k}_{\parallel}}e^{-{k}_{\parallel}|z-z'|},
\end{eqnarray}
with ${k}_{\parallel}^2=k^2_{x}+k^2_{y}$, the Green's function becomes
\begin{eqnarray}
  G(\mathbf{r},\mathbf{r}')
  &=&
  {4\pi}\int\frac{\text{d}^2{k}_{\parallel}}{(2\pi)^2}
  e^{\mathtt{i}\mathbf{k}_{\parallel}
  \cdot(\mathbf{r}_{\parallel}-\mathbf{r}'_{\parallel})}
  g(z,z';{k}_{\parallel}),
  \label{green_fn}
\end{eqnarray}
where $\mathbf{r}_{\parallel}=(x,y)$ is the 2D coordinates,
$\mathbf{k}_{\parallel}=(k_x,k_y)$ is the 2D $k$ vector, and
\begin{eqnarray}
  g(z,z';{k}_{\parallel})
  =
  \frac{1}{2\kappa{k}_{\parallel}}e^{-{k}_{\parallel}|z-z'|}
\end{eqnarray}
is the reduced Green's function. The differential equation for the reduced Green's
function can be found as
\begin{eqnarray}
  \kappa\left[{k}_{\parallel}^2-\frac{\partial^2}{\partial{z}^2}\right]
  g(z,z';{k}_{\parallel})=\delta(z-z').
\end{eqnarray}
By these two integrals over the infinitesimal segment near $z=z'$
\begin{eqnarray}
  \int^{z'+0}_{z'-0}\text{d}z
  \left[{k}_{\parallel}^2-\frac{\partial^2}{\partial{z}^2}\right]g(z,z';{k}_{\parallel})
  =
  \int^{z'+0}_{z'-0}\text{d}z\frac{\delta(z-z')}{\kappa},\n
\end{eqnarray}
\begin{eqnarray}
  &&\int^{z'+0}_{z'-0}\text{d}z\;z
  \left[{k}_{\parallel}^2-\frac{\partial^2}{\partial{z}^2}\right]g(z,z';{k}_{\parallel})\n
  &&=
  \int^{z'+0}_{z'-0}\text{d}z\;z\frac{\delta(z-z')}{\kappa},
\end{eqnarray}
the first equation gives
\begin{eqnarray}
  -\frac{\partial{g}}{\partial{z}}\Big\vert_{z=z'+0}
  +\frac{\partial{g}}{\partial{z}}\Big\vert_{z=z'-0}=\frac{1}{\kappa},
  \label{condition1}
\end{eqnarray}
with $0$ indicates an infinitesimal value, and the second equation becomes
\begin{eqnarray}
  -\int^{z'+0}_{z'-0}\text{d}z\;z\frac{\partial^2}{\partial{z}^2}
  g(z,z';{k}_{\parallel})=\frac{z'}{\kappa},
\end{eqnarray}
and then becomes
\begin{eqnarray}
  -z\frac{\partial{g}}{\partial{z}}\Big\vert^{z'+0}_{z'-0}
  +\int^{z'+0}_{z'-0}\text{d}z\frac{\partial{g}}{\partial{z}}
  =
  \frac{z'}{\kappa}.
\end{eqnarray}
By using Eq.~(\ref{condition1}), it is found that 
\begin{eqnarray}
  g\big\vert_{z=z'+0}=g\big\vert_{z=z'-0}.
  \label{condition2}
\end{eqnarray}
The equation indicates the continuity condition for the solution of the Poisson equation.
Eq.~(\ref{condition1}) and Eq.~(\ref{condition2}) indicate that the $g$ is continuous and
$\partial{g}/\partial{z}$ is discontinuous at $z=z'$.

\subsection{Rytova-Keldysh potential in parallel bilayers}

The parallel bilayers separate three regimes. Two are uniform dielectric materials $z<0$,
$z>d$. The regime in the middle $0<z<d$ is an empty space. The Green's function is written
as
\begin{eqnarray}
  &&z\leq{0}: 
  \hskip0.5ex
  \kappa_{1}\nabla^2G(\mathbf{r},\mathbf{r}')
  =-4\pi[\delta(\mathbf{r}-\mathbf{r}')+\sigma_{1}\delta(z)],\\
  &&0<z<d: 
  \hskip0.5ex
  \kappa_{0}\nabla^2G(\mathbf{r},\mathbf{r}')=-4\pi\delta(\mathbf{r}-\mathbf{r}'),\\
  &&z\geq{d}: 
  \hskip0.5ex
  \kappa_{2}\nabla^2G(\mathbf{r},\mathbf{r}')
  =-4\pi[\delta(\mathbf{r}-\mathbf{r}')+\sigma_{2}\delta(z-d)],\n
\end{eqnarray}
where $\sigma_{1}$, $\sigma_{2}$ are the surface bound charges on the layers, $\kappa_1$,
$\kappa_2$ are the dielectric constants of the uniform dielectric materials, and
$\kappa_{0}$ is the dielectric constant for the empty space. The surface-bound charge can
be related to the polarization of the layer by
\begin{eqnarray}
  \bsym{\nabla}_{\parallel}\cdot\mathbf{P}_{\parallel}=-\sigma,
\end{eqnarray}
where $\bsym{\nabla}_{\parallel}=(\partial/\partial{x},\partial/\partial{y})$ and
$\mathbf{P}_{\parallel}=(P_{x},P_{y})$ are the divergence and the polarization on the 2D
space. The polarization can be related to the electric field by
\begin{eqnarray}
  \mathbf{P}_{\parallel}
  =
  \chi_{\text{2D}}\mathbf{E}_{\parallel},
\end{eqnarray}
with $\mathbf{E}_{\parallel}=(E_x,E_y)$ the electric field on the two-dimensional space
and $\chi_{\text{2D}}$ is the electric susceptibility (or polarizability) of the layer.
Since $\mathbf{E}_{\parallel}=-\bsym{\nabla}_{\parallel}\Phi$, we can derive the 2D
Poisson equation,
\begin{eqnarray}
  \chi_{1}\nabla^2_{\parallel}\Phi(\mathbf{r})\big\vert_{z=0}=\sigma_{1},
  \hskip2ex
  \chi_{2}\nabla^2_{\parallel}\Phi(\mathbf{r})\big\vert_{z=d}=\sigma_{2},
\end{eqnarray}
where $\chi_{1}$ and $\chi_{2}$ are the electric susceptibilities of the corresponding
layers, the Green's function can be rewritten as
\begin{eqnarray}
  &&z\leq{0}:\n
  &&\kappa_{1}\left[\nabla^2
  +\frac{4\pi}{\kappa_{1}}\chi_{1}\delta(z)\nabla^2_{\parallel}\right]
  G(\mathbf{r},\mathbf{r}')
  =
  -4\pi\delta(\mathbf{r}-\mathbf{r}'),\\
  &&0<z<d: 
  \hskip1ex
  \kappa_{0}\nabla^2G(\mathbf{r},\mathbf{r}')
  =-4\pi\delta(\mathbf{r}-\mathbf{r}'),\\
  &&z\geq{d}:\n
  &&\kappa_{2}\left[\nabla^2
  +\frac{4\pi}{\kappa_{2}}\chi_{2}\delta(z-d)\nabla^2_{\parallel}\right]
  G(\mathbf{r},\mathbf{r}')
  =
  -4\pi\delta(\mathbf{r}-\mathbf{r}').\n
\end{eqnarray}
By using Eq.~(\ref{green_fn}), the Poisson equation for the reduced Green's function can
be written as
\begin{eqnarray}
  &&z\leq{0}:\hskip1ex
  \kappa_{1}\left[{k}_{\parallel}^2-\frac{\partial^2}{\partial{z}^2}
  +\frac{4\pi\chi_{1}}{\kappa_{1}}{k}_{\parallel}^2\delta(z)\right]g
  =
  \delta(z-z'),\n
  &&\\
  &&0<z<d:\hskip1ex
  \kappa_{0}\left[{k}_{\parallel}^2-\frac{\partial^2}{\partial{z}^2}\right]g
  =
  \delta(z-z'),\\
  &&z\geq{d}:\n
  &&\kappa_{2}\left[{k}_{\parallel}^2-\frac{\partial^2}{\partial{z}^2}
  +\frac{4\pi\chi_{2}}{\kappa_{2}}{k}_{\parallel}^2\delta(z-d)\right]g
  =
  \delta(z-z').
\end{eqnarray}
Assuming that the test charge is at $z'=0$, the boundary conditions can be derived 
\begin{eqnarray}
  -\kappa_{0}\frac{\partial}{\partial{z}}g\Big\vert_{z=+0}
  +\kappa_{1}\frac{\partial}{\partial{z}}g\Big\vert_{z=-0}
  =
  1-4\pi\chi_{1}{k}_{\parallel}^2g\big\vert_{z=0},\n
\end{eqnarray}
\begin{eqnarray}
  -\kappa_{2}\frac{\partial}{\partial{z}}g\Big\vert_{z=d+0}
  +\kappa_{0}\frac{\partial}{\partial{z}}g\Big\vert_{z=d-0}
  =
  -4\pi\chi_{2}{k}_{\parallel}^2g\big\vert_{z=d}.\n
\end{eqnarray}
The general solution of Green's function is assumed as
\begin{eqnarray}
  z\leq{0}:&\hskip2ex& g=Ae^{{k}_{\parallel} z},\\
  0<z<d:&\hskip2ex& g=Be^{{k}_{\parallel} z}+Ce^{-{k}_{\parallel} z},\\
  z\geq{d}:&\hskip2ex& g=De^{-{k}_{\parallel} z},
\end{eqnarray}
By using the continuity condition in Eq.~(\ref{condition2}) and the boundary contions, the
four coefficients can be connected to four coupled linear equations,
\begin{eqnarray}
  A=B+C,\hskip2ex
  Be^{{k}_{\parallel} d}+Ce^{-{k}_{\parallel} d}=De^{-{k}_{\parallel} d}.
\end{eqnarray}
\begin{eqnarray}
  \left(\kappa_{1}{k}_{\parallel}+2\xi_{1}{k}_{\parallel}^2\right)A
  -\kappa_{0}{k}_{\parallel}B+\kappa_{0}{k}_{\parallel}C=1,
\end{eqnarray}
\begin{eqnarray}
  &&\kappa_{0}{k}_{\parallel}\left(Be^{{k}_{\parallel}{d}}-Ce^{-{k}_{\parallel}{d}}\right)\n
  &&+\left(\kappa_{2}{k}_{\parallel}+4\pi\chi_{2}{k}_{\parallel}^2\right)
  De^{-{k}_{\parallel}{d}}
  =
  0.
\end{eqnarray}
By using the first equation $A=B+C$ in other linear equations to replace the coefficient
$A$, we can reduce one of the four variables. We get the three coupled equations,
\begin{eqnarray}
  Be^{{k}_{\parallel} d}+Ce^{-{k}_{\parallel} d}-De^{-{k}_{\parallel} d}=0.
\end{eqnarray}
\begin{eqnarray}
  (\lambda_{1}-\kappa_{0})B+(\lambda_{1}+\kappa_{0})C=1/{k}_{\parallel},
\end{eqnarray}
\begin{eqnarray}
  \kappa_{0}Be^{{k}_{\parallel}{d}}-\kappa_{0}Ce^{-{k}_{\parallel}{d}}
  +\lambda_{2} De^{-{k}_{\parallel}{d}}=0.
\end{eqnarray}
with $\lambda_{1}=\kappa_{1}+4\pi\chi_{1}{k}_{\parallel}$ and
$\lambda_{2}=\kappa_{2}+4\pi\chi_{2}{k}_{\parallel}$. These equations can be rewritten as
the matrix form,
\begin{eqnarray}
  \begin{bmatrix}
    \lambda_{1}-\kappa_{0} & \lambda_{1}+\kappa_{0} & 0\\
    e^{{k}_{\parallel} d} & e^{-{k}_{\parallel} d} & -e^{-{k}_{\parallel} d}\\
    \kappa_{0}e^{{k}_{\parallel}{d}} & -\kappa_{0}e^{-{k}_{\parallel}{d}} 
    & \lambda_{2}e^{-{k}_{\parallel}{d}}
  \end{bmatrix}
  \begin{bmatrix}
    B\\ C\\ D
  \end{bmatrix}
  =
  \begin{bmatrix}
    1/{k}_{\parallel}\\ 0\\ 0
  \end{bmatrix},\hskip3ex
\end{eqnarray}
By Cramer's rule, these coefficients can be solved as
\begin{eqnarray}
  B
  &=&
  \frac{1}{\textit{Det}}
  \begin{vmatrix}
    1/{k}_{\parallel} & \lambda_{1}+\kappa_{0} & 0\\
    0 & e^{-{k}_{\parallel} d} & -e^{-{k}_{\parallel} d}\\
    0 & -\kappa_{0} e^{-{k}_{\parallel}{d}} & \lambda_{2}e^{-{k}_{\parallel}{d}}
  \end{vmatrix}\n
  &=&
  \frac{(\lambda_{2}-\kappa_{0})e^{-2{k}_{\parallel}{d}}}{{k}_{\parallel}\textit{Det}},
\end{eqnarray}
\begin{eqnarray}
  C
  &=&
  \frac{1}{\textit{Det}}
  \begin{vmatrix}
    \lambda_{1}-\kappa_{0} & 1/{k}_{\parallel} & 0\\
    e^{{k}_{\parallel} d} & 0 & -e^{-{k}_{\parallel} d}\\
    \kappa_{0}e^{{k}_{\parallel}{d}} & 0 & \lambda_{2}e^{-{k}_{\parallel}{d}}
  \end{vmatrix}
  =
  -\frac{\lambda_2+\kappa_{0}}{{k}_{\parallel}\textit{Det}},\hskip3ex
\end{eqnarray}
\begin{eqnarray}
  D
  &=&
  \frac{1}{\textit{Det}}
  \begin{vmatrix}
    \lambda_{1}-\kappa_{0} & \lambda_{1}+\kappa_{0} & 1/{k}_{\parallel}\\
    e^{{k}_{\parallel} d} & e^{-{k}_{\parallel} d} & 0\\
    \kappa_{0}e^{{k}_{\parallel}{d}} & -\kappa_{0}e^{-{k}_{\parallel}{d}} & 0
  \end{vmatrix}
  =
  -\frac{2\kappa_{0}}{{k}_{\parallel}\textit{Det}},\hskip3ex
\end{eqnarray}
\begin{eqnarray}
  \textit{Det}
  &=&
  \begin{vmatrix}
    \lambda_{1}-\kappa_{0} & \lambda_{1}+\kappa_{0} & 0\\
    e^{{k}_{\parallel} d} & e^{-{k}_{\parallel} d} & -e^{-{k}_{\parallel} d}\\
    \kappa_{0}e^{{k}_{\parallel}{d}} & -\kappa_{0}e^{-{k}_{\parallel}{d}} 
    & \lambda_{2}e^{-{k}_{\parallel}{d}}
  \end{vmatrix}\n
  &=&
  (\lambda_{1}-\kappa_{0})(\lambda_{2}-\kappa_{0})e^{-2{k}_{\parallel}{d}}\n
  &&-(\lambda_{1}+\kappa_{0})(\lambda_{2}+\kappa_{0}).
\end{eqnarray}
Therefore, the intralayer and interlayer Rytova-Keldysh potentials can be found as
\begin{eqnarray}
  &&W_{11}({k}_{\parallel})
  =
  4\pi g\big\vert_{z=0,\;z'=0}\n
  &&=
  \frac{(4\pi/{k}_{\parallel})\left[(\lambda_{2}+\kappa_{0})e^{{k}_{\parallel}{d}}
  -(\lambda_{2}-\kappa_{0})e^{-{k}_{\parallel}{d}}\right]}
  {(\lambda_{1}+\kappa_{0})(\lambda_{2}+\kappa_{0})e^{{k}_{\parallel}{d}}
  -(\lambda_{1}-\kappa_{0})(\lambda_{2}-\kappa_{0})e^{-{k}_{\parallel}{d}}},\n
\end{eqnarray}
\begin{eqnarray}
  &&W_{21}({k}_{\parallel})
  =
  4\pi g\big\vert_{z=d,\;z'=0}\n
  &&=
  \frac{8\pi\kappa_{0}/{k}_{\parallel}}
  {(\lambda_{1}+\kappa_{0})(\lambda_{2}+\kappa_{0})e^{{k}_{\parallel}{d}}
  -(\lambda_{1}-\kappa_{0})(\lambda_{2}-\kappa_{0})e^{-{k}_{\parallel}{d}}}.\n
\end{eqnarray}
By replacing $q=k_{\parallel}$ and using dielectric functions, these potentials can be
rewritten as
\begin{eqnarray}
  W_{11}(q)
  =
  \frac{2\pi}{\epsilon_{11}(q)q},
  \hskip1ex
  W_{21}(q)
  =
  \frac{2\pi}{\epsilon_{21}(q)q},
\end{eqnarray}
where the intralayer and interlayer dielectric functions are written as 
\begin{eqnarray}
  \epsilon_{11}(q)
  &=&
  \frac{\kappa_{0}\epsilon_{21}(q)}
  {\left(\frac{\kappa_{2}+\kappa_{0}}{2}+r_{2}q\right)e^{q{d}}
  -\left(\frac{\kappa_{2}-\kappa_{0}}{2}
  +r_{2}q\right)e^{-q{d}}},\n
\end{eqnarray}
\begin{eqnarray}
  \epsilon_{21}(q)
  &=&
  \left(\frac{\kappa_{1}+\kappa_{0}}{2}+r_{1}q\right)
  \left(\frac{\kappa_{2}+\kappa_{0}}{2}+r_{2}q\right)
  \frac{e^{q{d}}}{\kappa_{0}}\n
  &&-
  \left(\frac{\kappa_{1}-\kappa_{0}}{2}+r_{1}q\right)
  \left(\frac{\kappa_{2}-\kappa_{0}}{2}+r_{2}q\right)
  \frac{e^{-q{d}}}{\kappa_{0}},\n
\end{eqnarray}
with $r_1=2\pi\chi_{1}$ and $r_2=2\pi\chi_{2}$ being the screening lengths. By assigning
$\kappa_1=\kappa_2$, the modified Rytova-Keldysh potentials in Eq.~(\ref{intra_1}),
Eq.~(\ref{intra_2}), and Eq.~(\ref{inter}) can be derived.

\begin{widetext}

\section{Matrix elements of exciton Hamiltonian\label{sec:matrix_element}}

Matrix elements of the exciton Hamiltonian matrix and overlap matrix expanded by the basis
functions in Sec.~\ref{ssec:exciton_wavefunction} are derived. By using the variational
method, the exciton wavefunction can be solved from the matrix eigenvalue equation
\begin{eqnarray}
  \sum_{l'_{\text{e}}l'_{\text{h}},b}
  \bar{\mathcal{H}}_{(l_{\text{e}}l_{\text{h}},{a},\mathbf{G})
  (l'_{\text{e}}l'_{\text{h}},{b},\mathbf{G}'),\mathbf{K}}
  C_{(l'_{\text{e}}l'_{\text{h}},{b},\mathbf{G}'),I\mathbf{K}}
  =
  \varepsilon_{\text{X},I\mathbf{K}}\sum_{b}
  \mathcal{O}_{(l_{\text{e}}l_{\text{h}},a,\mathbf{G})
  (l_{\text{e}}l_{\text{h}},b,\mathbf{G}')}
  C_{(l_{\text{e}}l_{\text{h}},{b},\mathbf{G}'),I\mathbf{K}}.
\end{eqnarray}
The exciton Hamiltonian matrix elements are given by
\begin{eqnarray}
  \bar{\mathcal{H}}_{(l_{\text{e}}l_{\text{h}},{a},\mathbf{G})
  (l'_{\text{e}}l'_{\text{h}},{b},\mathbf{G}'),\mathbf{K}}
  &=&
  \delta_{l_{\text{e}},l'_{\text{e}}}\delta_{l_{\text{h}},l'_{\text{h}}}
  \int\psi^*_{\mathbf{G}}(\mathbf{R})\phi^*_{l_{\text{e}}l_{\text{h}},a}(\mathbf{r})
  \tilde{\mathcal{H}}_{l_{\text{e}}l_{\text{h}}}(\mathbf{R},\mathbf{r})
  \psi_{\mathbf{G}'}(\mathbf{R})\phi_{l_{\text{e}}l_{\text{h}},b}(\mathbf{r})
  \text{d}^2R\text{d}^2r\n
  &&+
  (1-\delta_{l_{\text{e}},l'_{\text{e}}})\delta_{l_{\text{h}},l'_{\text{h}}}
  \int\psi^*_{\mathbf{G}}(\mathbf{R})\phi^*_{l_{\text{e}}l_{\text{h}},a}(\mathbf{r})
  \tilde{\mathcal{T}}_{\text{e},(l_{\text{e}}l_\text{h})(l'_{\text{e}}l_\text{h})}
  (\mathbf{R},\mathbf{r})
  \psi_{\mathbf{G}'}(\mathbf{R})\phi_{l'_{\text{e}}l_{\text{h}},b}(\mathbf{r})
  \text{d}^2R\text{d}^2r\n
  &&+\delta_{l_{\text{e}},l'_{\text{e}}}(1-\delta_{l_{\text{h}},l'_{\text{h}}})
  \int\psi^*_{\mathbf{G}}(\mathbf{R})\phi^*_{l_{\text{e}}l_{\text{h}},a}(\mathbf{r})
  \tilde{\mathcal{T}}_{\text{h},(l_\text{e}l_{\text{h}})(l_\text{e}l'_{\text{h}})}
  (\mathbf{R},\mathbf{r})
  \psi_{\mathbf{G}'}(\mathbf{R})\phi_{l_{\text{e}}l'_{\text{h}},b}(\mathbf{r})
  \text{d}^2R\text{d}^2r,
\end{eqnarray}
with
\begin{eqnarray}
  \tilde{\mathcal{H}}_{l_{\text{e}}l_{\text{h}}}(\mathbf{R},\mathbf{r})
  =
  e^{-\mathtt{i}\mathbf{K}\cdot\mathbf{R}}
  \exp\left(-{\mathtt{i}\bsym{\kappa}_{l_\text{e}l_\text{h}}\cdot\mathbf{r}}\right)
  \mathcal{H}_{l_{\text{e}}l_{\text{h}}}(\mathbf{R},\mathbf{r})
  \exp\left({\mathtt{i}\bsym{\kappa}_{l_\text{e}l_\text{h}}\cdot\mathbf{r}}\right)
  e^{\mathtt{i}\mathbf{K}\cdot\mathbf{R}},
\end{eqnarray}
\begin{eqnarray}
  \tilde{\mathcal{T}}_{\text{e},(1l_\text{h})(2l_\text{h})}
  (\mathbf{R},\mathbf{r})
  =
  e^{-\mathtt{i}\mathbf{K}\cdot\mathbf{R}}
  \exp\left(-{\mathtt{i}\bsym{\kappa}_{1l_\text{h}}\cdot\mathbf{r}}\right)
  \mathcal{T}_{\text{e}}(\mathbf{R},\mathbf{r})
  \exp\left({\mathtt{i}\bsym{\kappa}_{2l_\text{h}}\cdot\mathbf{r}}\right)
  e^{\mathtt{i}\mathbf{K}\cdot\mathbf{R}},
\end{eqnarray}
\begin{eqnarray}
  \tilde{\mathcal{T}}_{\text{e},(2l_\text{h})(1l_\text{h})}
  (\mathbf{R},\mathbf{r})
  =
  e^{-\mathtt{i}\mathbf{K}\cdot\mathbf{R}}
  \exp\left(-{\mathtt{i}\bsym{\kappa}_{2l_\text{h}}\cdot\mathbf{r}}\right)
  \mathcal{T}^{*}_{\text{e}}(\mathbf{R},\mathbf{r})
  \exp\left({\mathtt{i}\bsym{\kappa}_{1l_\text{h}}\cdot\mathbf{r}}\right)
  e^{\mathtt{i}\mathbf{K}\cdot\mathbf{R}},
\end{eqnarray}
\begin{eqnarray}
  \tilde{\mathcal{T}}_{\text{h},(l_\text{e}1)(l_\text{e}2)}(\mathbf{R},\mathbf{r})
  =
  e^{-\mathtt{i}\mathbf{K}\cdot\mathbf{R}}
  \exp\left(-{\mathtt{i}\bsym{\kappa}_{l_\text{e}1}\cdot\mathbf{r}}\right)
  \mathcal{T}_{\text{h}}(\mathbf{R},\mathbf{r})
  \exp\left({\mathtt{i}\bsym{\kappa}_{l_\text{e}2}\cdot\mathbf{r}}\right)
  e^{\mathtt{i}\mathbf{K}\cdot\mathbf{R}},
\end{eqnarray}
\begin{eqnarray}
  \tilde{\mathcal{T}}_{\text{h},(l_\text{e}2)(l_\text{e}1)}
  (\mathbf{R},\mathbf{r})
  =
  e^{-\mathtt{i}\mathbf{K}\cdot\mathbf{R}}
  \exp\left(-{\mathtt{i}\bsym{\kappa}_{l_\text{e}2}\cdot\mathbf{r}}\right)
  \mathcal{T}^*_{\text{h}}(\mathbf{R},\mathbf{r})
  \exp\left({\mathtt{i}\bsym{\kappa}_{l_\text{e}1}\cdot\mathbf{r}}\right)
  e^{\mathtt{i}\mathbf{K}\cdot\mathbf{R}}.
\end{eqnarray}
The overlap matrix elements are given by
\begin{eqnarray}
  \bar{\mathcal{O}}_{(l_{\text{e}}l_{\text{h}},{a},\mathbf{G})
  (l_{\text{e}}l_{\text{h}},{b},\mathbf{G}')}
  =
  \int\psi^*_{\mathbf{G}}(\mathbf{R})\phi^*_{l_{\text{e}}l_{\text{h}},a}(\mathbf{r})
  \psi_{\mathbf{G}'}(\mathbf{R})\phi_{l_{\text{e}}l_{\text{h}},b}(\mathbf{r})
  \text{d}^2R\text{d}^2r.
\end{eqnarray}
The diagonal Hamiltonian matrix elements are given by
\begin{eqnarray}
  \bar{\mathcal{H}}_{(l_{\text{e}}l_{\text{h}},{a},\mathbf{G})
  (l_{\text{e}}l_{\text{h}},{b},\mathbf{G}'),\mathbf{K}}
  &=&
  \delta_{\mathbf{G},\mathbf{G}'}o_{ab}\left[\Delta_{l_{\text{e}}l_{\text{h}}}
  +\frac{|\mathbf{K}-\mathbf{G}-\bsym{\mathcal{K}}_{l_\text{e}l_\text{h}}|^2}
  {2m_{\text{X},l_{\text{e}}l_{\text{h}}}}\right]
  +\delta_{\mathbf{G},\mathbf{G}'}\langle{a}|\left[-
  \frac{\nabla^2}{2\mu_{\text{X},l_{\text{e}}l_{\text{h}}}}
  -W_{l_{\text{e}}l_{\text{h}}}(r)\right]|{b}\rangle\n
  &&+\langle{a,\mathbf{G}}|\mathcal{V}_{l_{\text{e}}l_{\text{h}}}(\mathbf{R},\mathbf{r})
  |{b,\mathbf{G}'}\rangle,
\end{eqnarray}
\begin{eqnarray}
  -\frac{1}{2\mu_{\text{X},l_{\text{e}}l_{\text{h}}}}\langle{a}|\nabla^2|{b}\rangle
  &=&
  -\frac{1}{2\mu_{\text{X},l_{\text{e}}l_{\text{h}}}}
  \frac{(N_a+N_b-1)!}{\left(\mathcal{Z}_{l_{\text{e}}l_{\text{h}},a}
  +\mathcal{Z}_{l_{\text{e}}l_{\text{h}},b}\right)^{N_a+N_b}}
  \Bigg\{(1-\delta_{N_b,1})
  \frac{\left[(N_b-1)^2-L_b^2\right]\left(\mathcal{Z}_{l_{\text{e}}l_{\text{h}},a}
  +\mathcal{Z}_{l_{\text{e}}l_{\text{h}},b}\right)^2}
  {(N_a+N_b-1)(N_a+N_b-2)}\n
  &&-\frac{\left[(2N_b-1)\mathcal{Z}_{l_{\text{e}}l_{\text{h}},b}\right]
  \left(\mathcal{Z}_{l_{\text{e}}l_{\text{h}},a}
  +\mathcal{Z}_{l_{\text{e}}l_{\text{h}},b}\right)}{(N_a+N_b-1)}
  +\mathcal{Z}^2_{l_{\text{e}}l_{\text{h}},b}\Bigg\},
\end{eqnarray}
\begin{eqnarray}
  \langle{a}|W_{l_{\text{e}}l_{\text{h}}}(\mathbf{r})|{b}\rangle
  &=&
  \frac{\delta_{L_a,L_b}}{(2\pi)^2}\int^{\infty}_0
  \tilde{\mathcal{R}}_{N_a+N_b-1,0}(\mathcal{Z}_{l_{\text{e}}l_{\text{h}},a}
  +\mathcal{Z}_{l_{\text{e}}l_{\text{h}},b},k)
  \tilde{W}_{l_{\text{e}}l_{\text{h}}}(k)k\text{d}k,
  \label{potential_integral}
\end{eqnarray}
\begin{eqnarray}
  \langle{a,\mathbf{G}}|\mathcal{V}_{l_{\text{e}}l_{\text{h}}}(\mathbf{R},\mathbf{r})
  |{b,\mathbf{G}'}\rangle
  &=&
  \sum_{j=1,3,5}
  \Big\{
  V_{l_{\text{e}}}\exp\left[\mathtt{i}(-1)^{l_{\text{e}}}\psi\right]
  \delta(\mathbf{G}-\mathbf{G}'+\bsym{g}_{j})
  \tilde{\rho}_{(l_{\text{e}}l_{\text{h}},a)(l_{\text{e}}l_{\text{h}},b)}
  (-\gamma_{\text{h},l_{\text{h}}}\bsym{g}_{j})\n
  &&+V_{l_{\text{e}}}\exp\left[-\mathtt{i}(-1)^{l_{\text{e}}}\psi\right]
  \delta(\mathbf{G}-\mathbf{G}'-\bsym{g}_{j})
  \tilde{\rho}_{(l_{\text{e}}l_{\text{h}},a)(l_{\text{e}}l_{\text{h}},b)}
  (\gamma_{\text{h},l_{\text{h}}}\bsym{g}_{j})\n
  &&-V_{l_{\text{h}}}\exp\left[\mathtt{i}(-1)^{l_{\text{h}}}\psi\right]
  \delta(\mathbf{G}-\mathbf{G}'+\bsym{g}_{j})
  \tilde{\rho}_{(l_{\text{e}}l_{\text{h}},a)(l_{\text{e}}l_{\text{h}},b)}
  (\gamma_{\text{e},l_{\text{e}}}\bsym{g}_{j})\n
  &&-V_{l_{\text{h}}}\exp\left[-\mathtt{i}(-1)^{l_{\text{h}}}\psi\right]
  \delta(\mathbf{G}-\mathbf{G}'-\bsym{g}_{j})
  \tilde{\rho}_{(l_{\text{e}}l_{\text{h}},a)(l_{\text{e}}l_{\text{h}},b)}
  (-\gamma_{\text{e},l_{\text{e}}}\bsym{g}_{j})\Big\},
\end{eqnarray}
with
\begin{eqnarray} 
  \tilde{\rho}_{(l_{\text{e}}l_{\text{h}},a)(l'_{\text{e}}l'_{\text{h}},b)}(\mathbf{k})
  &=&
  \frac{e^{-\mathtt{i}(L_a-L_b)\varphi_{\mathbf{k}}}}{2\pi}
  \tilde{\mathcal{R}}_{N_a+N_b-1,L_b-L_a}(\mathcal{Z}_{l_{\text{e}}l_{\text{h}},a}
  +\mathcal{Z}_{l'_{\text{e}}l'_{\text{h}},b},k),
\end{eqnarray}
where the radial function in momentum space can be obtained by the generating formula
\begin{eqnarray}
  \tilde{\mathcal{R}}_{N,L}(\mathcal{Z},k)
  &=&
  \frac{2\pi(-\mathtt{i})^{N}}{k^{N+1}}
  \frac{\text{d}^N}{\text{d}z^N}
  \frac{\left(z-\mathtt{i}(L/|L|)\sqrt{1-z^2}\right)^{|L|}}{\sqrt{1-z^2}}
  \Bigg\vert_{z=\mathtt{i}{\mathcal{Z}}/{k}}.
  \label{Rk_formula}
\end{eqnarray}
The nondiagonal Hamiltonian matrix elements can be rewritten as
\begin{eqnarray}
  \bar{\mathcal{H}}_{(l_{\text{e}}l_{\text{h}},a,\mathbf{G})
  (l'_{\text{e}}l'_{\text{h}},b,\mathbf{G}'),\mathbf{K}}
  &=&
  w_{\text{e}}\delta_{l_{\text{h}},l'_{\text{h}}}(1-\delta_{l_{\text{e}},l'_{\text{e}}})
  \Big[\delta(\mathbf{G}-\mathbf{G}')
  \tilde{\rho}_{(l_{\text{e}}l_{\text{h}},a)(l'_{\text{e}}l_{\text{h}},b)}
  \big((l_{\text{e}}-l'_{\text{e}})(\gamma_{\text{h},l_\text{h}}\bsym{\kappa}_{3}
  +\Delta_{\text{e}}\bsym{\kappa}_{l_\text{h}})\big)\n
  &&+\delta\big(\mathbf{G}-\mathbf{G}'-(l_{\text{e}}-l'_{\text{e}})\bsym{g}_{1}\big)
  \tilde{\rho}_{(l_{\text{e}}l_{\text{h}},a)(l'_{\text{e}}l_{\text{h}},b)}
  \big((l_{\text{e}}-l'_{\text{e}})(\gamma_{\text{h},l_\text{h}}\bsym{\kappa}_{1}
  +\Delta_{\text{e}}\bsym{\kappa}_{l_\text{h}})\big)\n
  &&+\delta\big(\mathbf{G}-\mathbf{G}'-(l_{\text{e}}-l'_{\text{e}})\bsym{g}_{2}\big)
  \tilde{\rho}_{(l_{\text{e}}l_{\text{h}},a)(l'_{\text{e}}l_{\text{h}},b)}
  \big((l_{\text{e}}-l'_{\text{e}})(\gamma_{\text{h},l_\text{h}}\bsym{\kappa}_{5}
  +\Delta_{\text{e}}\bsym{\kappa}_{l_\text{h}})\big)\Big]\n
  &&+w_{\text{h}}\delta_{l_{\text{e}},l'_{\text{e}}}(1-\delta_{l_{\text{h}},l'_{\text{h}}})
  \Big[\delta(\mathbf{G}-\mathbf{G}')
  \tilde{\rho}_{(l_{\text{e}}l_{\text{h}},a)(l_{\text{e}}l'_{\text{h}},b)}
  \big((l_{\text{h}}-l'_{\text{h}})(\gamma_{\text{e},l_\text{e}}\bsym{\kappa}_{3}
  +\Delta_{\text{h}}\bsym{\kappa}_{l_\text{e}})\big)\n
  &&+\delta\big(\mathbf{G}-\mathbf{G}'+(l_{\text{h}}-l'_{\text{h}})\bsym{g}_{1}\big)
  \tilde{\rho}_{(l_{\text{e}}l_{\text{h}},a)(l_{\text{e}}l'_{\text{h}},b)}
  \big((l_{\text{h}}-l'_{\text{h}})(\gamma_{\text{e},l_\text{e}}\bsym{\kappa}_{1}
  +\Delta_{\text{h}}\bsym{\kappa}_{l_\text{e}})\big)\n
  &&+\delta\big(\mathbf{G}-\mathbf{G}'+(l_{\text{h}}-l'_{\text{h}})\bsym{g}_{2}\big)
  \tilde{\rho}_{(l_{\text{e}}l_{\text{h}},a)(l_{\text{e}}l'_{\text{h}},b)}
  \big((l_{\text{h}}-l'_{\text{h}})(\gamma_{\text{e},l_\text{e}}\bsym{\kappa}_{5}
  +\Delta_{\text{h}}\bsym{\kappa}_{l_\text{e}})\big)\Big],
\end{eqnarray}
with $\Delta_{\text{e}}=\gamma_{\text{e},2}-\gamma_{\text{e},1}$,
$\Delta_{\text{h}}=\gamma_{\text{h},2}-\gamma_{\text{h},1}$. The overlap matrix element is
given by
\begin{eqnarray}
  \bar{\mathcal{O}}_{(l_{\text{e}}l_{\text{h}},{a},\mathbf{G})
  (l_{\text{e}}l_{\text{h}},{b},\mathbf{G}')}
  &=&
  \delta(\mathbf{G}-\mathbf{G}')
  \delta_{L_a,L_b}\frac{(N_a+N_b-1)!}
  {(\mathcal{Z}_{l_{\text{e}}l_{\text{h}},a}
  +\mathcal{Z}_{l_{\text{e}}l_{\text{h}},b})^{N_a+N_b}}.
\end{eqnarray}
A more detailed derivation of matrix elements of the exciton Hamiltonian by using STOs as
the basis function can be found in Ref.~\onlinecite{mypaper0, mypaper1}.

\end{widetext}

\end{document}